\documentclass[10pt,english,twoside]{article}
\usepackage{authblk}

\usepackage{graphicx}
\usepackage{epstopdf,epsfig}
\usepackage[table]{xcolor}
\usepackage{pgfplots}
\pgfplotsset{compat=1.3}
\usetikzlibrary{external}
\usepgfplotslibrary{groupplots}
\tikzset{external/mode=graphics if exists}
\usepackage{relsize}
\def\solidthick{\protect\rule[2pt]{10.pt}{1pt}}

\definecolor{my_dark_blue}{rgb}{0.0,0.0,0.5}
\definecolor{my_blue}{rgb}{0.0,0.4,0.8}
\definecolor{my_cyan}{rgb}{0.13,0.82,0.82}
\definecolor{my_lightblue}{rgb}{0.66,0.76,0.83}
\definecolor{my_petrol}{rgb}{0.19,0.55,0.55}
\definecolor{my_green}{rgb}{0.0,0.6,0.51}
\definecolor{my_green2}{rgb}{0.16,0.75,0.47}
\definecolor{my_brgreen}{rgb}{0.67,0.84,0.19}
\definecolor{my_orange}{rgb}{0.9,0.55,0.0}
\definecolor{my_red}{rgb}{1.0,0.0,0.0}
\definecolor{my_wine_red}{rgb}{0.83,0.0,0.35}
\definecolor{my_dark_red}{rgb}{0.63,0.12,0.10}
\definecolor{my_purple}{rgb}{0.4,0.0,0.6}
\definecolor{col_qhuntneg}{rgb}{0.1,0.3,0.5}
\definecolor{col_qhuntpos}{rgb}{0.71,0.27,0.15} %[179 69 39]/255
\definecolor{mygray}{rgb}{0.5,0.5,0.5}
\definecolor{col_gray}{rgb}{0.63,0.63,0.63}
\definecolor{col_gray2}{rgb}{0.3,0.3,0.3}
\definecolor{bg-color2}{rgb}{0.156,0.173,0.204}
\definecolor{col_nosym}{rgb}{0.00,0.09,0.31}
\definecolor{col_yrefl}{rgb}{0.1059,0.6,0.0784}
\definecolor{col_yzrefl}{rgb}{0.76,0.00,0.28}

\definecolor{col_20000}{rgb}{0.3,0.3,0.3}
\definecolor{col_20004}{rgb}{0.3,0.3,0.3}
\definecolor{col_16001}{rgb}{0.3,0.3,0.3}
\definecolor{col_18001}{rgb}{0.3,0.3,0.3}
\definecolor{col_20001}{rgb}{0.3,0.3,0.3}
\definecolor{col_22001}{rgb}{0.3,0.3,0.3}
\definecolor{col_24001}{rgb}{0.3,0.3,0.3}
\definecolor{col_26001}{rgb}{0.3,0.3,0.3}
\definecolor{col_28001}{rgb}{0.3,0.3,0.3}
\definecolor{col_30001}{rgb}{0.3,0.3,0.3}
\definecolor{col_32001}{rgb}{0.3,0.3,0.3}
\definecolor{col_16002}{rgb}{0.3,0.3,0.3}
\definecolor{col_18002}{rgb}{0.3,0.3,0.3}
\definecolor{col_20002}{rgb}{0.3,0.3,0.3}
\definecolor{col_22002}{rgb}{0.3,0.3,0.3}
\definecolor{col_24002}{rgb}{0.3,0.3,0.3}
\definecolor{col_26002}{rgb}{0.3,0.3,0.3}
\definecolor{col_28002}{rgb}{0.3,0.3,0.3}
\definecolor{col_30002}{rgb}{0.3,0.3,0.3}
\definecolor{col_32002}{rgb}{0.3,0.3,0.3}
\definecolor{col_1150}{rgb}{0.3,0.3,0.3}
\definecolor{col_1200}{rgb}{0.3,0.3,0.3}

\definecolor{col1_matlab}{rgb}{0.000,0.447,0.741}
\definecolor{col2_matlab}{rgb}{0.850,0.325,0.098}
\definecolor{col3_matlab}{rgb}{0.929,0.694,0.125}
\definecolor{col4_matlab}{rgb}{0.494,0.184,0.556}
\definecolor{col5_matlab}{rgb}{0.466,0.674,0.188}
\definecolor{col6_matlab}{rgb}{0.301,0.745,0.933}
\definecolor{col7_matlab}{rgb}{0.635,0.078,0.184}

\pgfplotsset{tikzPlotsDefault/.style=
    {hide axis,scale only axis,
     height=0pt,width=0pt,
     colorbar right,
     colorbar style={
             anchor=west,
             major tick length=0.05cm,
             ylabel near ticks,
             label style={font=\footnotesize},
             ylabel shift={-0.05cm}
             }
     }
}

\pgfplotsset{colorbar style={
        anchor=west,
        major tick length=0.05cm,
        ylabel near ticks,
        label style={font=\small}
        }
}

\pgfplotsset{
    colormap={bluewhite}{
        rgb255=(  0, 39, 93)
        rgb255=(252,252,252)
    },
}
\pgfplotsset{
    colormap={greenwhite}{
        rgb255=( 17, 93, 10)
        rgb255=(252,252,252)
    },
}
\pgfplotsset{
    colormap={redwhite}{
        rgb255=(174,  0, 71)
        rgb255=(252,252,252)
    },
}

\usepgfplotslibrary{colormaps}
\usepackage{amsmath,amssymb,amsfonts,scalerel}

\usepackage{mathrsfs}
\usepackage{mathtools}
\usepackage{accents}
\usepackage{gensymb}
\usepackage{textcomp}
\usepackage{siunitx}
\usepackage{trimclip}
\usepackage{ar}
\usepackage[colorlinks,linkcolor=blue,citecolor=blue,urlcolor=blue]{hyperref}
\usepackage[
a4paper,
headheight=1.5\baselineskip,
top=25mm,
bottom=25mm,
inner=25mm,
outer=25mm,
heightrounded=true]
{geometry}

\usepackage{array}
\newcolumntype{H}{>{\setbox0=\hbox\bgroup}c<{\egroup}@{}}
\usepackage{arydshln} % allows for dashed hlines
\usepackage[round,sort&compress]{natbib}

\let\doi\undefined
\usepackage{doi}

\usepackage[normalem]{ulem}
\usepackage{etex}
\newcommand{\revisions}[1]{}

\title{On the role of turbulent large-scale streaks in generating sediment ridges}

\author[1]{Markus Scherer \footnote{Email address for correspondence: \href{mailto:markus.scherer@kit.edu}{markus.scherer@kit.edu}}}
\author[1]{Markus Uhlmann}
\author[2]{Genta Kawahara}
\affil[1]{\small Institute for Water and Environment -- Numerical Fluid Mechanics Group,
        Karlsruhe Institute of Technology, 76131 Karlsruhe, Germany}
\affil[2]{\small Graduate School of Engineering Science, Osaka University,
        1-3 Machikaneyama, Toyonaka, Osaka 560-8531, Japan}
\vspace{-1.3cm}
\date{\small (Dated: \today)}

\title{Chaotic and time-periodic edge states in square duct flow}

\newcommand\abs[1]{\ensuremath{\vert {#1} \vert}}
\newcommand\norm[1]{\ensuremath{\vert\vert {#1} \vert\vert}}
\newcommand\Lnorm[2]{\ensuremath{\left\lVert {#1} \right\rVert_{#2}}}

\newcommand\avg[1]{\ensuremath{\langle {#1} \rangle}}
\newcommand\xavg[1]{\ensuremath{\langle {#1} \rangle_{x}}}

\newcommand\tavg[1]{\ensuremath{\langle {#1} \rangle_{t}}}
\newcommand\xtavg[1]{\ensuremath{\langle {#1} \rangle_{xt}}}

\newcommand\volavg[1]{\ensuremath{\langle {#1} \rangle_{V}}}
\newcommand\casenosymshort[1]{\ensuremath{\mathrm{CA}{#1}_{2\pi}}}
\newcommand\casenosymlong[1]{\ensuremath{\mathrm{CA}{#1}_{4\pi}}}
\newcommand\caseyrefl[1]{\ensuremath{\mathrm{PO}{#1}_y}}
\newcommand\caseyzrefl[1]{\ensuremath{\mathrm{PO}{#1}_{yz}}}

\newcommand\viz{\textit{viz.}}
\newcommand\realn{\ensuremath{\mathbb{R}}}
\newcommand\natn{\ensuremath{\mathbb{N}}}

\newcommand\Id{\ensuremath{\mathrm{Id}}}

\newcommand\xvecshift{\ensuremath{\boldsymbol{s}}}
\newcommand\xshift{\ensuremath{s_x}}
\newcommand\xshiftop{\ensuremath{l_x}}

\newcommand\alp{\ensuremath{\alpha}}

\newcommand\tper{\ensuremath{T}}
\newcommand\tperpre{\ensuremath{T}_{pre}}

\newcommand\edgemfd{\ensuremath{\mathscr{M}}}
\newcommand\Ekin{\ensuremath{{E}}}

\newcommand\Epert{\ensuremath{{E}_{3D}}}
\newcommand\Epertrms{\ensuremath{{E}_{rms}}}
\newcommand\Epertrmsl{\ensuremath{{E}_{rms}^l}}
\newcommand\Epertrmst{\ensuremath{{E}_{rms}^t}}
\newcommand\Ep{\ensuremath{{E_p}}}
\newcommand\Estreak{\ensuremath{{E_{streak}}}}
\newcommand\Erolls{\ensuremath{{E_{rolls}}}}
\newcommand\Ewave{\ensuremath{{E_{wave}}}}
\newcommand\Einp{\ensuremath{{I}}}
\newcommand\Einplam{\ensuremath{{I}_{lam}}}
\newcommand\Diss{\ensuremath{{D}}}
\newcommand\Disslam{\ensuremath{{D}_{lam}}}
\newcommand\I{\ensuremath{\mathrm{Id}_{\triangle}}}

\newcommand\hf{\ensuremath{{H}}}

\newcommand\idmap{\ensuremath{e}}
\newcommand\yrefl{\ensuremath{\mathrm{\mathbf{Z}}_y}}
\newcommand\zrefl{\ensuremath{\mathrm{\mathbf{Z}}_z}}
\newcommand\pihalfrot{\ensuremath{\mathrm{\mathbf{R}}_{\pi/2}}}
\newcommand\pirot{\ensuremath{\mathrm{\mathbf{R}}_\pi}}
\newcommand\pithrhalfrot{\ensuremath{\mathrm{\mathbf{R}}_{3\pi/2}}}
\newcommand\yshrefl{\ensuremath{\mathrm{\mathbf{S}}_y}}
\newcommand\zshrefl{\ensuremath{\mathrm{\mathbf{S}}_z}}
\newcommand\pishrot{\ensuremath{\mathrm{\mathbf{S}}_\pi}}
\newcommand\yzdiag{\ensuremath{\mathrm{\mathbf{D}}_{yz}}}
\newcommand\zydiag{\ensuremath{\mathrm{\mathbf{D}}_{zy}}}

\newcommand\avec{\ensuremath{\boldsymbol{a}}}

\newcommand\xvec{\ensuremath{\boldsymbol{x}}}

\newcommand\dom{\ensuremath{\Omega}}
\newcommand\domperp{\ensuremath{\Omega_{\perp}}}

\newcommand\gradientperp{\ensuremath{\nabla_{\perp}}}

\newcommand\tbulk{\ensuremath{T_b}}

\newcommand\rhof{\ensuremath{\rho}} %  fluid density
\newcommand\fnu{\ensuremath{\nu}} % fluid viscosity (kin.)
\newcommand\hmean{\ensuremath{H_f}}

\newcommand\Lx{\ensuremath{L_x}}
\newcommand\Ly{\ensuremath{L_y}}
\newcommand\Lz{\ensuremath{L_z}}
\newcommand\uvec{\ensuremath{\boldsymbol{u}}}

\newcommand\uveclam{\ensuremath{\boldsymbol{u}_{lam}}}
\newcommand\uplvec{\ensuremath{\boldsymbol{u}_p}}
\newcommand\upl{\ensuremath{u_p}}
\newcommand\vpl{\ensuremath{v_p}}
\newcommand\wpl{\ensuremath{w_p}}

\newcommand\uvecL{\ensuremath{\boldsymbol{u}^L}}
\newcommand\uvecH{\ensuremath{\boldsymbol{u}^H}}
\newcommand\uvecxt{\ensuremath{\xtavg{\boldsymbol{u}}}}

\newcommand\uvecfour{\ensuremath{\hat{\boldsymbol{u}}}}

\newcommand\usec{\ensuremath{u_{\perp}}}

\newcommand\ubulk{\ensuremath{u_b}}
\newcommand\utau{\ensuremath{u_\tau}}

\newcommand\dpdx{\ensuremath{f_e}}
\newcommand\dpdxvec{\ensuremath{\boldsymbol{f}_e}}
\newcommand\dpdxlam{\ensuremath{f_{e,lam}}}

\newcommand\tauw{\ensuremath{\overline{\tau}_w}}
\newcommand\tauwi{\ensuremath{\tau_{w,i}}}
\newcommand\tauwilam{\ensuremath{\tau_{w,i,lam}}}
\newcommand\tauwxy{\ensuremath{\tau_{xy}}}
\newcommand\tauwxz{\ensuremath{\tau_{xz}}}

\newcommand\psixt{\ensuremath{\xtavg{\psi}}}

\newcommand\omvec{\ensuremath{\boldsymbol{\omega}}}
\newcommand\omx{\ensuremath{\omega_{x}}}
\newcommand\omy{\ensuremath{\omega_{y}}}
\newcommand\omz{\ensuremath{\omega_{z}}}
\newcommand\omxxt{\ensuremath{\xtavg{\omega_{f,x}}}}
\newcommand{\rhobed}{\ensuremath{\rhobed}} % bed density
\newcommand\Reb{\ensuremath{Re_b}}    % Bulk Reynolds number
\newcommand\Ret{\ensuremath{Re_\tau}} % friction Reynolds number
\newcommand\Retau{\ensuremath{Re_\tau}} % same, but i use them interchangeably
\newcommand\Nx{\ensuremath{N_x}}
\newcommand\Ny{\ensuremath{N_y}}
\newcommand\Nz{\ensuremath{N_z}}
\newcommand\deltat{\ensuremath{\Delta t}} % the other

\newcommand\deltaxplus{\ensuremath{\Delta x^+}}
\newcommand\deltayplus{\ensuremath{\Delta y^+}}
\newcommand\deltazplus{\ensuremath{\Delta z^+}}

\newcommand\tsep{\ensuremath{\delta t}}

\begin{document}

\maketitle

%===================================================================
% short version of previous works:
%
\newcommand{\TI}{TI03} % (Toh & Itano, JFM, 2003)
\newcommand{\KVSE}{KVSE13} % (Kreilos et al., JFM, 2013)
\newcommand{\SUK}{SUK24} % (Scherer, Uhlmann & Kawahara, IOP, 2024)

%===================================================================
\begin{abstract}
We analyse the dynamics within the stability boundary between laminar and turbulent square duct flow with the aid of an edge-tracking algorithm. As for the circular pipe, the edge state turns out to be a chaotic attractor within the edge if the flow is not constrained to a symmetric subspace.
The chaotic edge state dynamics is characterised by a sequence of alternating quiescent phases and regularly occurring bursting episodes. These latter reflect the different stages of the well-known streak-vortex interaction in near-wall turbulence: The edge states feature most of the time a single streak with a number of flanking quasi-streamwise vortices attached to one of the four surrounding walls.
The initially straight streak undergoes the classical linear instability and eventually breaks in an intense bursting event due to the action of the quasi-streamwise vortices. At the same time, the downstream vortices give rise to a new generation of low-speed streaks at one of the neighbouring walls, thereby causing the turbulent activity to `switch' from one wall to the other.
When restricting the edge dynamics to a single or twofold mirror-symmetric subspace, on the other hand, the outlined bursting and wall-switching episodes become self-recurrent in time. These edge states thus represent the first periodic orbits found in the square duct. In contrast to the chaotic edge states in the non-symmetric case, though, the imposed symmetries enforce analogue bursting cycles to simultaneously appear at two parallel opposing walls in a mirror-symmetric configuration.
Both localisation of the turbulent activity to one or two walls and wall-switching are shown to be a common phenomenon in low Reynolds number duct turbulence. We therefore argue that the marginally turbulent trajectories transiently visit the identified edge states during these episodes, so that the edge states become actively involved in the turbulent dynamics.

\end{abstract}

%===================================================================
  \section{Introduction}\label{sec:intro}
  Turbulent flows in straight pipes of various cross-sections are of utmost relevance for many industrial processes.
For circular pipes, the turbulent dynamics both in the transitional and the fully-developed regime are nowadays quite well understood \citep{Smits_McKeon_Marusic_2011,Avila_Barkley_Hof_2023}. 
Ducts with a rectangular cross-section, on the other hand, have attained much less attention, despite their relevance in many technical applications such as air ventilation systems. In contrast to the circular case, the rotational symmetry w.r.t. the centreline is broken by the four surrounding walls, a two-dimensional mean velocity field with a distinct mean secondary flow of Prandtl's second kind is the consequence \citep{Bradshaw_1987}.
Although of comparably low amplitude (usually about $\mathcal{O}(1\%)$ of the bulk velocity $\ubulk$), these mean secondary currents are of significance for the mass, momentum and heat transfer between the near-wall zones, the corner regions and the duct core \citep{Demuren_Rodi_1984,Modesti_Pirozzoli_2022}.
In a streamwise- and time-averaged sense, the turbulence-induced secondary flow is generated by the inhomogeneity and anisotropy of the Reynolds-stress tensor across the duct cross-section \citep{Speziale_1982}.
Hence, most previous experimental studies \citep{Brundrett_Baines_1964,Gessner_1973}
and direct numerical simulations (DNS) \citep{Gavrilakis_1992,Gavrilakis_2019,%
Modesti_Pirozzoli_Orlandi_Grasso_2018,Pirozzoli_2018} laid great emphasis on a detailed analysis of the mean momentum, vorticity and energy budgets \citep{Nikora_Roy_2012}.

While such investigations of the mean flow equations are insightful, for instance, regarding the scaling of the mean secondary flow pattern and intensity with the Reynolds number, they provide little information about the underlying physical mechanisms and the involved coherent flow structures.
\citet{Uhlmann_al_2007} and \citet{Pinelli_al_2010} therefore performed detailed coherent structure eduction studies for a range of marginal to moderate Reynolds numbers, assessing the role of individual streaks and quasi-streamwise vortices in the generation of the characteristic mean secondary flow pattern.
\citet{Uhlmann_al_2007} showed that for sufficiently high Reynolds numbers $\Reb=\ubulk\hf/\fnu$ (where $\hf$ is the duct's half side-length and $\fnu$ is the kinematic viscosity), each of the four bounding walls accommodates one or more streaks with flanking quasi-streamwise vortices that undergo the classical self-sustained near-wall regeneration cycle \citep{Hamilton_Kim_Waleffe_1995,Waleffe_1997,Hall_Sherwin_2010}.
However, the quasi-streamwise vortices next to the corner regions are limited in their spatial mobility by the two adjacent walls \citep{Kawahara_Kamada_2000}, so that they get essentially locked in their cross-sectional position and
leave a clear footprint in the mean streamwise vorticity $\omxxt$.
The resulting preferential organisation of small-scale vortices in certain regions of the cross-section has direct implications for the large-scale secondary flow, as that is coupled to the small scale dynamics by a Poisson equation $\nabla^2\psixt=-\omxxt$; with $\psixt$ indicating the mean secondary flow streamfunction \citep{Pinelli_al_2010}.

%==============================================================
% insert figure:
\begin{figure}%[tp]
    \centering
    \includegraphics[width=0.95\linewidth]
    {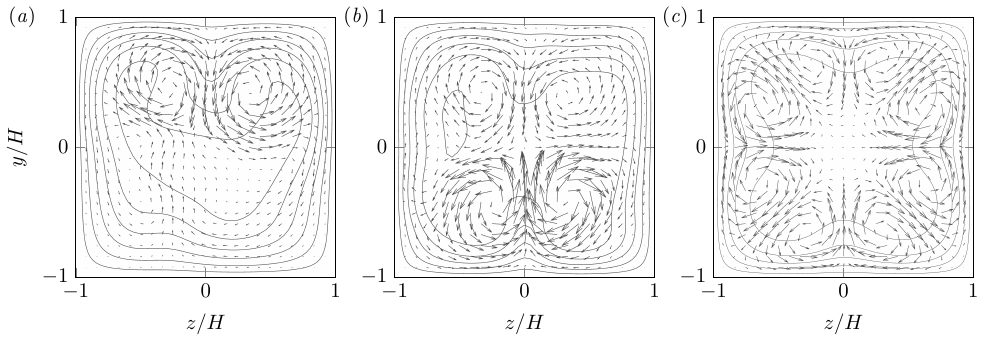}
    \caption
    {
    %\MS{TBA}
    (\textit{a,b}) Selected instantaneous streamwise-averaged velocity fields $\xavg{\uvec}$
    % representative for (\textit{a}) a single-wall "two-vortex" and (\textit{b}) a two-wall "four-vortex state".
    %The flow states are
    taken from a long-time turbulent trajectory of $\mathcal{O}(10^4)$ bulk time units length at a bulk Reynolds number $\Reb=1150$ and a streamwise domain length $\Lx=2\pi\hmean$.
    (\textit{c}) Long-time average of the velocity field $\xtavg{\uvec}$ over the full trajectory.
    In all panels, solid lines represent isolines of $\avg{u}$ between $0$ and $1.5$ times the bulk velocity $\ubulk$, with increment $0.25$.
    Intensity and orientation of the secondary flow field $(\avg{v},\avg{w})^T$
    is indicated by the vector plot.
    The arrows in (\textit{c}) are scaled by a factor of two compared to those in (\textit{a,b}) for better visibility.
    For a complete definition of the velocity field and the averaging operators, see \S~\ref{sec:theory}.
    }
\label{fig:uvw_xavg_snaps_closedDuctref}
\end{figure}

%==============================================================
%
At marginal Reynolds numbers $\Reb$ just large enough to sustain turbulence, on the other hand, the typical diameter of the quasi-streamwise vortices (which naturally scale in wall units) is with $\mathcal{O}(0.1\hf)$ so large that independent streak-vortex pairs cannot always be sustained on all four walls simultaneously \citep{Uhlmann_al_2007}.
Instead, only some of the four surrounding walls feature a significant turbulent/vortical activity, while turbulent fluctuations are weak or even entirely absent near the remaining ones during such episodes.
In the remainder of this study, we will occasionally refer to these two situations as `active' and `non-active' walls, respectively.
\autoref{fig:uvw_xavg_snaps_closedDuctref}\textit{a,b} present the streamwise-averaged velocity field for two such flow states extracted from a long-time turbulent trajectory at $\Reb=1150$, in which a single and two opposing walls are `active' in the above defined sense, respectively.
Such localised states are transient in time and a `switching' of the vortical activity to other walls takes place intermittently in intervals of $\mathcal{O}(100)$ bulk time units length \citep{Uhlmann_al_2007}, though the underlying physical mechanisms are not entirely understood.
Only long-time averages over intervals much longer than the lifetime of the individual coherent structures and the characteristic `wall-switching' time-scale do recover the characteristic fourfold-symmetric eight-vortex mean secondary flow pattern under such marginal conditions (cf. \autoref{fig:uvw_xavg_snaps_closedDuctref}\textit{c}).
Both the `wall-switching' behaviour and the localisation of turbulent fluctuations to individual walls possess analogues in `minimal flow units' of plane channel flow \citep{Jimenez_Moin_1991}: For extended time periods, only one of the two channel walls hosts a streak with accompanying vortices, while the other one features essentially laminar dynamics until an intermittent switching event causes a reorganisation of the flow \citep{Neelavara_Duguet_Lusseyran_2017}.
Also, there is an equivalence between the discussed temporally intermittent `wall-switching' behaviour and a spatially intermittent `wall-switching' observed in DNS of extended domains like the $\Lx=160\pi\hf$ long duct studied by \citet{Sekimoto_2011}.
For a fixed moment in time, the dynamics along each wall is here characterised by a streamwise alternating pattern of `active' and `passive' regions, all together inducing the characteristic fourfold-symmetric eight-vortex mean secondary flow in the instantaneous streamwise average.

While the occurrence of spatially localised states and intermittent wall-switching events in low Reynolds number square duct turbulence is hence well documented, our understanding of the underlying physical mechanisms remains incomplete.
In this study, we analyse both phenomena from a dynamical systems' perspective on fluid turbulence \citep{Hopf_1948}, in which the time evolution of a turbulent system is understood as a trajectory wandering across the infinite-dimensional state space associated with the Navier-Stokes system.
Turbulent coherent structures are interpreted as physical incarnations of the least unstable invariant solutions in this state space \citep{Jimenez_1987,Gibson_Halcrow_Cvitanovic_2008}, which comprises stationary equilibria (EQ), travelling waves (TW, equilibria within a co-moving frame of reference) and periodic orbits (PO).
Together with their stable and unstable manifolds, these invariant solutions form the skeleton of the system's dynamics in state space: Chaotic turbulent trajectories visit their neighbourhood transiently and shadow their dynamics during that time, before they get repelled along their unstable manifolds \citep{Waleffe_2002}.
In the past decades, this conceptual picture has been extraordinarily fruitful for the understanding of transitional \citep{Kerswell_2005,Avila_Barkley_Hof_2023}
and sustained turbulence \citep{Kawahara_al_2012,Graham_2021} in a variety of different flow configurations.
Most prominently, equilibria, travelling waves and (relative) periodic orbits have been found
in plane Couette \citep{Nagata_1990,Kawahara_Kida_2001,Gibson_Halcrow_Cvitanovic_2008} and
plane Poiseuille flow \citep{Waleffe_2001,Waleffe_2003,Park_Shekar_Graham_2018},
asymptotic suction boundary layers \citep{Kreilos_Veble_Schneider_Eckhardt_2013,Khapko_al_2013} and
circular pipe flow \citep{Duguet_Willis_Kerswell_2008,Avila_al_2013,Budanur_al_2017}.
In square duct flows, on the other hand, only a handful of travelling waves are known to date:
The solution documented by \citet{Uhlmann_al_2010} generates an `eight-vortex' mean secondary flow pattern that bears a striking similarity with the long-time average in low-Reynolds number flows.
The `four-vortex' states of \citet{Wedin_Bottaro_Nagata_2009} and \citet{Okino_Nagata_Wedin_Bottaro_2010} and the `two-vortex' state of \citet{Okino_Nagata_2012}, in turn, are representative of situations in which not all of the four surrounding walls feature a significant vortical activity. In contrast to canonical flows such as
circular pipe or plane channel flows, however, no periodic orbits have been found so far in the square duct setting to the best of the authors' knowledge.

For all above discussed flow configurations including square duct flow \citep{Tatsumi_Yoshimura_1990}, the laminar flow state is stable w.r.t. infinitesimal disturbances for Reynolds numbers at which transition to turbulence usually sets in. Hence, finite-amplitude perturbations are required to trigger the transition to turbulence and the laminar flow takes the form of an equilibrium/fix point in state space, with a distinct basin of attraction \citep{Graham_2021}.
For transiently turbulent systems, the laminar state is complemented by a chaotic saddle to which nearby trajectories are attracted for a potentially long, but finite time horizon, before they finally experience a relaminarisation.
For sustained turbulence, two concurring state space pictures have been discussed: Turbulence might be further understood as a chaotic saddle, but with survival times of turbulent perturbations tending to infinity. Alternatively, turbulence is seen as a strange attractor from which trajectories cannot escape (cf. \citealp{Avila_Barkley_Hof_2023} for a recent review).

Irrespective of the chosen viewpoint, the state space of all above considered flows has been seen to feature an invariant manifold with co-dimension one \citep{Guckenheimer_Holmes_1983} that is conventionally termed the `edge' $\edgemfd$: Trajectories starting off from a point inside the edge will stay in the latter for all times, neither swinging up to a turbulent state nor decaying directly towards the laminar fix point. While the existence of the edge is consistent with both discussed state space pictures of turbulence, only in the presence of a turbulent attractor it becomes a separatrix of phase space in a strict sense \citep{Avila_Barkley_Hof_2023}.
Analysing the dynamics of such edge trajectories and identifying attracting sets within $\edgemfd$ (`edge states') has provided valuable insights into the transition processes to turbulence
\citep{Toh_Itano_2003,Mellibovsky_Meseguer_Schneider_Eckhardt_2009,%
Schneider_Eckhardt_Yorke_2007,Duguet_Schlatter_Henningson_2009}.
In this context, the nature of the detected edge states differs significantly among the different flows,
ranging from simple equilibria in Couette flow \citep{Schneider_al_2008}
or symmetric subspaces of circular pipe flow \citep{Avila_al_2013},
periodic orbits in plane channels \citep{Zammert_Eckhardt_2014,Rawat_Cossu_Rincon_2014}
and asymptotic suction boundary layers \citep{Kreilos_Veble_Schneider_Eckhardt_2013,%
Khapko_al_2013,Khapko_al_2014,Khapko_al_2016}
to chaotic edge states in the unconstrained circular pipe \citep{Duguet_Willis_Kerswell_2008}.

To the best of the authors' knowledge, only two groups have made efforts to analyse the edge for square duct flows:
\citet{Biau_Bottaro_2009} compared a rather short edge trajectory with a flow state that they had found to be optimal in terms of the maximum non-linear growth, without further discussing the edge dynamics or possible edge states. \citet[][in Japanese]{Okino_2014} and \citet{Gepner_Okino_Kawahara_2025} searched for edge states in a square duct under certain symmetry restrictions and in domains of streamwise periods $\Lx/\hf\in\{2\pi,4\pi,8\pi\}$.
For the three considered domains, a `domain-filling' TW, a chaotic (relative) attractor and a streamwise-localised TW were found to be edge states of the respective symmetry-reduced systems, respectively.
Especially the streamwise-localised solution bears a strong resemblance to localised `puff-like' solutions in the circular pipe \citep{Avila_al_2013}, with important implications for the understanding of localised turbulence in rectangular ducts \citep{Takeishi_al_2015}.

However, a comprehensive study addressing the edge dynamics and how they compare to the characteristics of chaotic low Reynolds number turbulence is lacking.
In this study, we therefore present such a detailed analysis of the edge in square duct flow for both, the symmetrically-unconstrained case and symmetric subspaces thereof.
We provide evidence that the edge states in the full state space are chaotic attractors within $\edgemfd$, whereas their counterparts in the symmetric subspaces turn out to be periodic orbits.
To the best of the authors' knowledge, these periodic solutions are the first of their kind and are hence considered to be of relevance for the understanding of transitional and marginally turbulent duct flows.
Specific implications for the understanding of transverse turbulence localisation and the process behind the `wall-switching' dynamics are discussed.

%-----------------------------------------------------------------
% Close with roadmap for the current manuscript
The manuscript is organised as follows:
In \S~\ref{sec:theory}, the considered physical system and the admissible symmetry groups are presented.
The essential features of the DNS code and the details of the edge-tracking algorithm are described in \S~\ref{sec:numerics}.
With the aid of these methods, the edge to turbulence in square duct flows is analysed in \S~\ref{sec:results}:
The edge dynamics in the full state space are addressed first,
before we turn to the edge states in symmetric subspaces.
In \S~\ref{sec:discussion}, the detected edge states are juxtaposed to states in marginally turbulent square duct flow
and their properties are discussed in comparison with edge states in other flows.
The manuscript ends with a summary of the relevant outcomes and an outlook on future work in \S~\ref{sec:conclusion}.

%===================================================================
  \section{Theoretical framework}\label{sec:theory}
  \subsection{Flow configuration}
In the present study, the flow of an incompressible Newtonian fluid in a straight, streamwise periodic duct of square cross-section with side lengths $2\hf$ is considered. No-slip boundary conditions are imposed along the four surrounding walls.
A Cartesian coordinate system with the origin located in the centre of the cross-section is adopted, with basis vectors pointing in the streamwise ($x$) and the two cross-stream, wall-normal directions ($y,z$). For the sake of clarity, we choose a terminology similar to that in plane Couette or Poiseuille flow and refer to $y$ and $z$ as the `vertical' and `spanwise' directions in the remainder of this work. Note that this designation is an arbitrary choice since the two cross-sectional directions are equivalent regarding the rotational symmetry obeyed by the governing equations.
In what follows, the velocity field is expressed as $\uvec(\xvec,t)=(u,v,w)^T$, with $u$, $v$ and $w$ indicating the fluid velocity along the $x$-, $y$- and $z$-direction, respectively. Analogously, the vorticity field is introduced as $\omvec(\xvec,t)\equiv\nabla\times\uvec=(\omx,\omy,\omz)^T$.

All flow variables are periodic in the streamwise direction, with fundamental period $\Lx=2\pi\hf/\alp$ and fundamental wavenumber $\alp$. If not otherwise stated, $\alp=1$ and thus $\Lx=2\pi\hf$ holds for the simulation results presented in this manuscript. Streamwise, volume and time averaging operators are consequently defined as
\begin{equation}
  \xavg{\bullet} \equiv \dfrac{\alp}{2\pi}\displaystyle\int\limits_0^{2\pi/\alp} \bullet \;\mathrm{d}x, \quad
  \volavg{\bullet} \equiv \dfrac{1}{\abs{\dom}}\displaystyle\int\limits_\dom \bullet \;\mathrm{d}\xvec, \quad
  \tavg{\bullet} \equiv \displaystyle\lim_{T\to\infty}\dfrac{1}{T}\displaystyle\int\limits_0^T \bullet \;\mathrm{d}t,
\end{equation}
where $\dom=[0,2\pi/\alp)\times[-\hf,\hf]\times[-\hf,\hf]$ denotes the physical domain.
In a similar fashion, the $L_2$-norm of a vector field $\avec$ is introduced as
\begin{equation}
  \Lnorm{\avec}{2}^2 \equiv \dfrac{1}{\abs{\dom}}\displaystyle\int\limits_\dom \avec\cdot\avec \;\mathrm{d}\xvec
                      = \volavg{\avec\cdot\avec}.
\end{equation}
%
% with volume $V=8\pi\hf^2/\alp$.
A time-dependent streamwise pressure gradient $\dpdxvec=(\dpdx(t),0,0)^T$ is imposed and adjusted at each time step to drive the flow at a constant mass flow rate $Q_m$, guaranteeing a fixed bulk velocity $\ubulk\equiv Q/(4\hf^2)$. As a consequence, both the perimeter-averaged wall shear stress
\begin{equation}
  \tauw(t)=\dfrac{\rho\fnu}{8\hf}
  \left(\,
  \displaystyle\int\limits_{z=-\hf}^{\hf}
    \left.\dfrac{\partial u}{\partial y}\right|_{y\in\{-\hf,\hf\}} \mathrm{d}z +
  \displaystyle\int\limits_{y=-\hf}^{\hf}
    \left.\dfrac{\partial u}{\partial z}\right|_{z\in\{-\hf,\hf\}} \mathrm{d}y
  \,\right)
\end{equation}
and the friction velocity $\utau(t)=\sqrt{\tauw(t)/\rhof}$ fluctuate in time. Variables non-dimensionalised with the time-averaged friction velocity $\tavg{\utau}$ and $\fnu$ are indicated as ${\bullet}^+$.
Based on this definition, all considered systems feature a domain clearly longer than the critical value required to sustain turbulence ($\Lx^+|_{min}\approx200$, cf. \citealp{Uhlmann_al_2007} and compare with table~\ref{tab:EdgeStates}) and are thus not minimal w.r.t. the buffer layer structures \citep{Jimenez_Moin_1991}.

For the considered case of a pressure-driven square duct flow, the overall kinetic energy budget simplifies to a balance between the energy input $\Einp$ and the dissipation rate $\Diss$ per unit mass, \viz
\begin{equation}
    \dfrac{\mathrm{d}\Ekin}{\mathrm{d}t}=\Einp-\Diss,
\end{equation}
where the individual contributions are defined as
\begin{equation}
    \Ekin=\dfrac{1}{2}\volavg{\uvec\cdot\uvec}, \hspace{4ex}
    \Einp=\volavg{\dpdxvec\cdot\uvec}, \hspace{4ex}
    \Diss=\fnu\volavg{\norm{\omvec}^2}.
\end{equation}
The laminar base flow $\uvec_{lam}=\left(u_{lam}(y,z),0,0\right)^T$ is defined as the solution of the following Poisson equation
\begin{equation}
  \fnu\gradientperp^2 u_{lam} = -\dpdxlam,
\end{equation}
where $\gradientperp^2=(\partial^2/\partial y^2 + \partial^2/\partial z^2)$ represents a two-dimensional Laplacian acting along the two cross-sectional directions only, and $\dpdxlam$ indicates the constant pressure gradient necessary to drive a laminar flow at identical mass flow rate $Q_m$. Perturbations of a given velocity field w.r.t. the laminar equilibrium are henceforth denoted as $\uplvec \equiv \uvec - \uvec_{lam}$.

\subsection{Symmetric subspaces}
Subject to the given boundary conditions, the Navier-Stokes equations admit several continuous and discrete symmetries. Streamwise periodicity induces a continuous translational shift symmetry in $x$, \viz
\begin{equation}
  \begin{array}{rlcl}
  \xshiftop(\xshift):& \left[u,v,w\right]\left(x,y,z\right) &\to&
              \left[u,v,w\right]\left(x+\xshift,y,z\right) \quad \forall \, \xshift\in\left[-\dfrac{\pi}{\alp},\dfrac{\pi}{\alp}\right),
  \end{array}
\end{equation}
allowing the existence of travelling waves (relative equilibria) and relative periodic orbits, where `relative' refers to invariance within a co-moving frame of reference. Unlike circular pipe flow, the rectangular cross-section cannot feature a continuous rotational symmetry. Instead, the symmetry group reduces to the discrete dihedral group $D_2=\{\idmap,\yrefl,\zrefl,\pirot\}$.
Here, $\idmap$ denotes the identity, $\yrefl,\zrefl$ are the two reflectional symmetries along the $y$- and $z$-direction, respectively, and $\pirot$ is the $\pi$-rotational symmetry around the $x$-axis, \viz
\begin{equation}
  \left.
\begin{array}{rlcl}
  \yrefl:&  \left[u,v,w\right]\left(x,y,z\right) &\to&
            \left[u,-v,w\right]\left(x,-y,z\right), \\[1.5ex]
  \zrefl:&  \left[u,v,w\right]\left(x,y,z\right) &\to&
            \left[u,v,-w\right]\left(x,y,-z\right), \\[1.5ex]
  \pirot:&  \left[u,v,w\right]\left(x,y,z\right) &\to&
            \left[u,-v,-w\right]\left(x,-y,-z\right).
\end{array}
\right. 
\end{equation}
For the special case of a square cross-section, the symmetry group increases to  $D_4=D_2\cup\{\yzdiag,\zydiag,\pihalfrot,\pithrhalfrot\}$.
In addition to the elements of $D_2$, $D_4$ hence contains reflectional symmetries w.r.t. the corner bisectors $y=z$ and $y=-z$ ($\yzdiag,\zydiag$) and rotational symmetries by counter-clockwise rotations of $\pi/2$ and $3\pi/2$ about the $x$-axis ($\pihalfrot,\pithrhalfrot$), \viz
\begin{equation}
  \left.
\begin{array}{rlcl}
  \yzdiag       :&  \left[u,v,w\right]\left(x,y,z\right) &\to&
                    \left[u,w,v\right]\left(x,z,y\right), \\[1.5ex]
  \zydiag       :&  \left[u,v,w\right]\left(x,y,z\right) &\to&
                    \left[u,-w,-v\right]\left(x,-z,-y\right),  \\[1.5ex]
  \pihalfrot    :&  \left[u,v,w\right]\left(x,y,z\right) &\to&
                    \left[u,-w,v\right]\left(x,-z,y\right), \\[1.5ex]
  \pithrhalfrot :&  \left[u,v,w\right]\left(x,y,z\right) &\to&
                    \left[u,w,-v\right]\left(x,z,-y\right).
\end{array}
\right. 
\end{equation}
Combining both streamwise and cross-sectional symmetries, it is readily seen that the governing equations are equivariant under the symmetry group $\Gamma=SO(2)_x\times D_4$. In the current work, the `shift-reflect' and `shift-rotate' elements of $\Gamma$ for fixed streamwise shifts $\Lx/2 = \pi/\alp$ are of specific interest and therefore deserve an explicit definition:
\begin{equation}
  \left.
\begin{array}{rlcl}
  \yshrefl:&  \left[u,v,w\right]\left(x,y,z\right) &\to&
              \left[u,-v,w\right]\left(x+\dfrac{\pi}{\alp},-y,z\right), \\[1.75ex]
  \zshrefl:&  \left[u,v,w\right]\left(x,y,z\right) &\to&
              \left[u,v,-w\right]\left(x+\dfrac{\pi}{\alp},y,-z\right), \\[1.75ex]
  \pishrot:&  \left[u,v,w\right]\left(x,y,z\right) &\to&
              \left[u,-v,-w\right]\left(x+\dfrac{\pi}{\alp},-y,-z\right).
\end{array}
\right. 
\end{equation}
%

%===================================================================
  \section{Numerical method}\label{sec:numerics}
  \subsection{Fluid solver}\label{sec:numerics_dnscode}
The pseudo-spectral DNS code developed by \citet{Pinelli_al_2010} is used to integrate a given flow state forward in time. Time stepping is realised using an incremental pressure projection scheme, in the context of which linear and non-linear contributions are treated with a semi-implicit Crank-Nicolson and a low-storage Runge-Kutta method \citep{Rai_Moin_1991,Verzicco_Orlandi_1996}, respectively. Throughout, simulation time steps are small enough to ensure that the CFL number stays below $0.3$.
The velocity and pressure fields are expanded in terms of truncated Fourier series in the periodic streamwise and Chebyshev polynomials in the two cross-stream directions.
While a grid of equispaced nodes is introduced in the streamwise direction, the standard Gauss–Chebyshev–Lobatto points are chosen as collocation points in the cross-stream directions.
Back and forth transformation between physical and Fourier space along the streamwise direction is performed with the aid of fast Fourier transforms and standard de-aliasing according to the $2/3$-rule is applied throughout. Fourier expansion of the fields along the streamwise direction allows to solve linear Poisson and Helmholtz equations for each streamwise wavenumber separately. The resulting two-dimensional systems only depend on the cross-sectional directions and can be efficiently solved with a fast-diagonalisation technique \citep{Haldenwang_1984}. The remaining non-linear contributions are solved in physical space.

For all edge-tracking simulations, a combination of $\Nx=\{64,128\}$ (for $\Lx/\hf=\{2\pi,4\pi\}$) Fourier modes and $\Ny=\Nz=129$ Chebyshev polynomials are chosen in the streamwise and the two cross-stream directions, respectively.
The resulting discretisation fulfils the resolution requirements for fully-turbulent trajectories at all considered Reynolds numbers, leading to inner-scaled grid spacings of $\deltaxplus\leq14$ and $\max(\deltayplus)=\max(\deltazplus)\leq 3.5$. Tests in which edge trajectories were integrated with less Chebyshev polynomials $\Ny=\Nz=\{65,97\}$ gave qualitatively similar results in that the trajectories converged to periodic edge states at comparable period. However, the coarser runs failed to resolve the bursting amplitudes of the higher-resolution data with sufficient precision, justifying the relatively high resolution in the two cross-stream directions. A lower resolution of $\Nx\times\Ny\times\Nz=48\times65\times65$ is merely used for the reference simulation at a marginal Reynolds number $\Reb=1150$, in which case this discretisation is sufficient to resolve all relevant flow scales.

\subsection{Edge-tracking}\label{sec:numerics_edgetracking}
Tracking trajectories along the edge is commonly done by a bisection between trajectories that approach the laminar fix point and those which leave the neighbourhood of the edge towards the turbulent attractor/saddle
\citep{Itano_Toh_2001,Toh_Itano_2003,Skufca_Yorke_Eckhardt_2006,Duguet_Willis_Kerswell_2008,%
Schneider_Eckhardt_Yorke_2007,Schneider_Eckhardt_2009}.
In the remainder, trajectories that return to the laminar state and those experiencing a rise in energy towards a turbulent state are denoted as $\uvecL$ and $\uvecH$, respectively.
While similar in concept, the different edge-tracking approaches feature some relevant differences concerning the states chosen on either side of the edge to perform the bisection. In this study, we apply a technique similar to that presented by \citet{Schneider_Eckhardt_Yorke_2007} to follow trajectories along the edge. In this approach, bisection is performed along a ray connecting a state $\uvecH$ that will tend to the turbulent `side' and the laminar equilibrium $\uveclam$.
As long as $\uvecH - \uveclam$ is not tangent to $\edgemfd$, the procedure per se allows to track edge trajectories for arbitrary time intervals. However, the positive Lyapunov exponents of the chaotic system cause any initially close trajectories to separate exponentially fast.
Hence, even if the bisection determines a flow state on the edge up to double precision, a `refinement step' after $\tsep=\mathcal{O}(100\tbulk)$ is unavoidable to find a new approximation for a state on $\edgemfd$ \citep{Toh_Itano_2003,Duguet_Willis_Kerswell_2008}. The finally obtained solution is therefore -- strictly speaking -- a piecewise-continuous approximation to a true edge trajectory, with small-amplitude discontinuities occurring at the refinement times $t_i$ \citep{Itano_Toh_2001}.

In the current work, the first bisection is initialised with a flow state $\uvec_0(\xvec,t_{0})$, randomly chosen from a turbulent trajectory at matching $\Reb$. The general procedure can then be summarised as follows: The $i$th bisection step iteratively seeks a parameter $\beta_i\in\realn$ ($\forall i\in\natn$) that rescales the velocity perturbation of a state $\uvecH_{i-1}(\xvec,t_{i-1})$ w.r.t. the laminar base flow at the beginning of the corresponding time interval $t_{i-1}$, \viz
\begin{equation}
  \left\{
  \begin{array}{rcll}
    \uvec_{\beta_i}(\xvec,t_{i-1})
    &\equiv& \uvec_{lam} + \beta_i \left(\uvecH_{i-1}(\xvec,t_{i-1}) -\uvec_{lam} \right)\quad & \text{at}\;\, t=t_{i-1} \\[0.8ex]
    \uvec_{\beta_i}(\xvec,t)
    &\equiv& \uvec_{\beta_i}(\xvec,t_{i-1})
    + \displaystyle\int\limits_{t'=t_{i-1}}^{t} \dfrac{\partial}{\partial t} \uvec \;\mathrm{d}t' \quad
    & \forall t\in(t_{i-1},t_{i})
  \end{array}
  \right..
  \label{eq:edge_tracking_def}
\end{equation}
In each of these shooting steps, the rescaled flow state is advanced in time according to
equation~\eqref{eq:edge_tracking_def} with the aid of the DNS code outlined in \S~\ref{sec:numerics_dnscode},
until the flow either relaminarises or becomes turbulent. 
As a measure for when a trajectory has left the edge, we use the root mean square (r.m.s.) 
of the kinetic energy of the non-constant streamwise Fourier modes, \viz
\begin{equation}
  \Epertrms\equiv\displaystyle\sum\limits_{j} [\Epert]_j^{{1}/{2}},
\end{equation}
with
\begin{equation}
  [\Epert]_j=
  \displaystyle\sum\limits_{i=1}^{\Nx-1} \;\;
  \displaystyle\iint\limits_{y,z=-\hf}^{\hf} \;
   [\uvecfour_{i}]_j[\uvecfour_{i}^{*}]_j
  \;\mathrm{d}z\mathrm{d}y \quad \forall j\in\{x,y,z\}.
\end{equation}
Therein, $\uvecfour_{i}$ and $\uvecfour_{i}^{*}$ indicate the $i$th Fourier coefficient and its complex conjugate, respectively, and $[\bullet]_j$ denotes the $j$th component of a scalar or vector field.
The trajectory is considered as deviated from $\edgemfd$ when $\Epertrms$ either falls below $\Epertrmsl\approx 2\cdot10^{-3}\ubulk$ or, following a steep energy rise, exceeds $\Epertrmst\approx 2\cdot10^{-1}\ubulk$.
Each bisection is continued until the relative error $\abs{\beta^H-\beta^L}/\abs{2\beta_i}$ falls below $10^{-12}$,
where $\beta_i$ represent the current approximation to the parameter $\beta$. The bracketing parameter values, eventually leading to a relaminarisation or a turbulent state, are denoted by $\beta^L$ and $\beta^H$, respectively.

Since the process is started with a turbulent flow state, the first parameter value $\beta_1$ is usually clearly lower than unity. For short enough refinement intervals, subsequent bisections start with a much better initial guess for a state within $\edgemfd$, so that all $\beta_i \;\forall i\geq2$ are very close to unity and, thus, the computed piecewise continuous curve is a good approximation to a true edge trajectory. In order to ensure this quality of the approximation, the trajectories $\uvecL$ and $\uvecH$ are advanced in time over $\tsep=300\tbulk$ (in single cases up to $\tsep=600\tbulk$), as long as it is guaranteed that the relative distance between the two bracketing states $\uvecL$ and $\uvecH$,
\begin{equation}
\varepsilon_{LH} \equiv \dfrac{\Lnorm{\uvecH(t_i + \tsep)-\uvecL(t_i + \tsep)}{2}}
                              {\Lnorm{\uvecH(t_i + \tsep)}{2} + \Lnorm{\uvecH(t_i + \tsep)}{2}},
\end{equation}
remains below a threshold of $10^{-4}$.
In case $\varepsilon_{LH}$ exceeds this limit, $\tsep$ is successively reduced until the criterion is fulfilled
\citep{Kreilos_Veble_Schneider_Eckhardt_2013}.

%===================================================================
  \section{Results}\label{sec:results}
  %==============================================================
% insert figure:
\begin{table}%[t]
\begin{center}
\begin{tabular}{l c c c c H c c c H}
\hline
\multicolumn{1}{c}{Case}&
\multicolumn{1}{c}{\Reb}&
\multicolumn{1}{c}{\Ret}&
\multicolumn{1}{c}{\Lx/\hf}&
\multicolumn{1}{c}{{\Lx}$^{+}$}&
\multicolumn{1}{H}{\AR}&
\multicolumn{1}{c}{\tper/\tbulk}&
\multicolumn{1}{c}{Symmetries}&
\multicolumn{1}{c}{Type}&
\multicolumn{1}{H}{Source}\\
\hline
{{\color{col_nosym} $\solidthick$}} \casenosymshort{2000} & $2000$ & $  88.7$ & $2\pi$ & $   557$ & $   1.0$ & - & - & CA & present\\[1.5pt] 
{{\color{col_nosym} $\solidthick$}} \casenosymlong{2000} & $2000$ & $  88.8$ & $4\pi$ & $  1116$ & $   1.0$ & - & - & CA & present\\[1.5pt] 
\hline
{{\color{col_yrefl} $\solidthick$}} \caseyrefl{1600} & $1600$ & $  81.4$ & $2\pi$ & $   511$ & $   1.0$ & 433.1 & \yrefl & PO & present\\[1.5pt] 
{{\color{col_yrefl} $\solidthick$}} \caseyrefl{2000} & $2000$ & $  90.5$ & $2\pi$ & $   569$ & $   1.0$ & 531.9 & \yrefl & PO & present\\[1.5pt] 
{{\color{col_yrefl} $\solidthick$}} \caseyrefl{2200} & $2200$ & $  94.8$ & $2\pi$ & $   595$ & $   1.0$ & 593.3 & \yrefl & PO & present\\[1.5pt] 
{{\color{col_yrefl} $\solidthick$}} \caseyrefl{2400} & $2400$ & $  98.8$ & $2\pi$ & $   621$ & $   1.0$ & 653.3 & \yrefl & PO & present\\[1.5pt] 
{{\color{col_yrefl} $\solidthick$}} \caseyrefl{3200} & $3200$ & $ 113.6$ & $2\pi$ & $   714$ & $   1.0$ & 3776.5 & \yrefl & PO & present\\[1.5pt] 
\hline
{{\color{col_yzrefl} $\solidthick$}} \caseyzrefl{1600} & $1600$ & $  86.7$ & $2\pi$ & $   545$ & $   1.0$ & 1397.0 & \yrefl,\zrefl & PO & present\\[1.5pt] 
{{\color{col_yzrefl} $\solidthick$}} \caseyzrefl{1800} & $1800$ & $  90.8$ & $2\pi$ & $   570$ & $   1.0$ & 403.7 & \yrefl,\zrefl & PO & present\\[1.5pt] 
{{\color{col_yzrefl} $\solidthick$}} \caseyzrefl{2000} & $2000$ & $  95.2$ & $2\pi$ & $   598$ & $   1.0$ & 401.9 & \yrefl,\zrefl & PO & present\\[1.5pt] 
{{\color{col_yzrefl} $\solidthick$}} \caseyzrefl{2200} & $2200$ & $  99.4$ & $2\pi$ & $   624$ & $   1.0$ & 432.8 & \yrefl,\zrefl & PO & present\\[1.5pt] 
{{\color{col_yzrefl} $\solidthick$}} \caseyzrefl{2400} & $2400$ & $ 103.4$ & $2\pi$ & $   650$ & $   1.0$ & 463.8 & \yrefl,\zrefl & PO & present\\[1.5pt] 
{{\color{col_yzrefl} $\solidthick$}} \caseyzrefl{2600} & $2600$ & $ 107.4$ & $2\pi$ & $   675$ & $   1.0$ & 492.7 & \yrefl,\zrefl & PO & present\\[1.5pt] 
{{\color{col_yzrefl} $\solidthick$}} \caseyzrefl{2800} & $2800$ & $ 111.2$ & $2\pi$ & $   699$ & $   1.0$ & 518.8 & \yrefl,\zrefl & PO & present\\[1.5pt] 
{{\color{col_yzrefl} $\solidthick$}} \caseyzrefl{3000} & $3000$ & $ 114.9$ & $2\pi$ & $   722$ & $   1.0$ & 542.4 & \yrefl,\zrefl & PO & present\\[1.5pt] 
{{\color{col_yzrefl} $\solidthick$}} \caseyzrefl{3200} & $3200$ & $ 118.7$ & $2\pi$ & $   746$ & $   1.0$ & 563.4 & \yrefl,\zrefl & PO & present\\[1.5pt] 
\hline
\end{tabular}
%
% ---- end of table generated in matlab ---
%

%
\caption{
 \label{tab:EdgeStates}
  Physical properties of the detected edge states. Here, $\Reb$ is the bulk, $\Retau$ the mean friction Reynolds number and $\Lx/\hf$, $\Lx^+$ indicate the outer and inner-scaled streamwise domain period, respectively.
  For time-periodic edge states, the period is given in bulk time units $\tbulk=\hf/\ubulk$.
  The last two columns list (potentially) imposed symmetries and provide information on the shape of the found edge state, respectively 
  (CA: chaotic attractor, PO: stable (relative) periodic orbit within $\edgemfd$).
}
\end{center}
\end{table}

%==============================================================

\subsection{Edge-state dynamics in the full state space}\label{subsec:nosymm_edge}
In a first step, edge trajectories are sought in the full phase space without restriction to any symmetric subspace. The characteristic dynamics will be analysed for trajectories in two domains of different streamwise period, $\Lx/\hf\in\{2\pi,4\pi\}$, at Reynolds number $\Reb=2000$. Trajectories at other Reynolds numbers revealed a qualitatively similar behaviour, so that the current findings are representative for a wider range of Reynolds numbers.
In the remainder of this work, the two considered cases will be referred to as \casenosymshort{2000} and \casenosymlong{2000}, respectively, and the properties of their edge states are summarised in \autoref{tab:EdgeStates}, together with the periodic edge states discussed in the subsequent section.

%
%==============================================================
% insert figure:
\begin{figure}%[tp]
    \centering
    \includegraphics[width=0.95\linewidth]
    {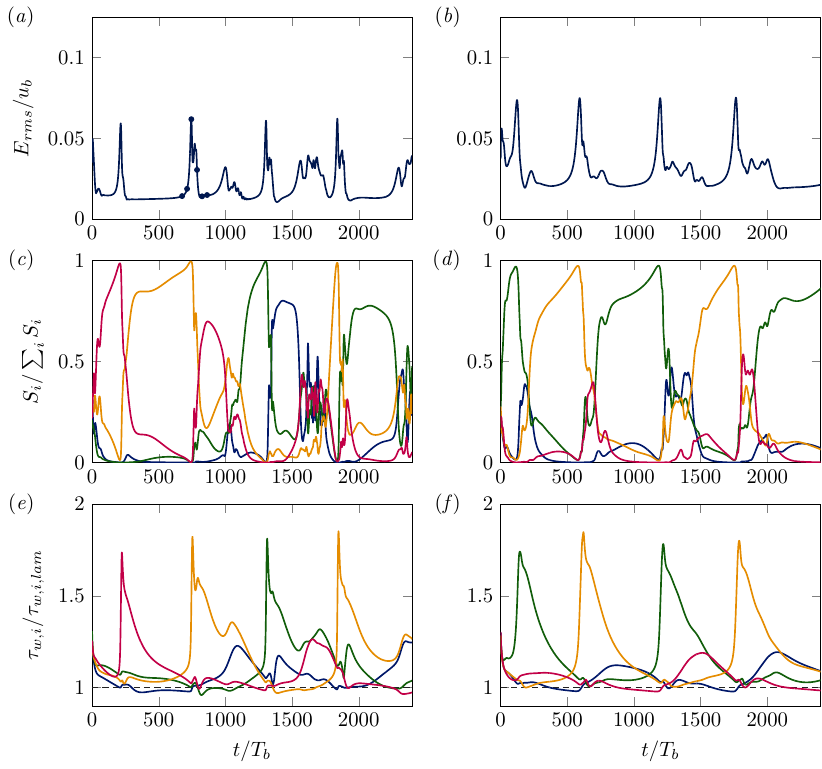}
    \caption
    {
    Time evolution of different flow measures for the non-symmetric edge trajectories
    of cases
    (\textit{a,c,e}) \casenosymshort{2000} and
    (\textit{b,d,f}) \casenosymlong{2000}.
    (\textit{a,b}) R.m.s. energy signal $\Epertrms(t)$, with solid nodes indicating snapshots that are visualised in subsequent figures.
    (\textit{c,d}) Individual contributions $S_i(t)$ of the triangular sectors to the overall mean streamwise enstrophy.
    (\textit{e,f}) Shear stress averaged along each of the four walls separately, $\tauwi(t)$, normalised by the corresponding laminar value $\tauwilam(t)$.
    In (\textit{c-f}), colour coding indicates the respective wall/sub-sector:
    $i={S}$ (blue),
    $i={E}$ (green),
    $i={N}$ (orange),
    $i={W}$ (red).
    }
\label{fig:E3drms_omxTri_tauWall_nosym}
\end{figure}

%==============================================================
In the absence of imposed symmetries, the edge states in the considered domains are relative chaotic attractors within $\edgemfd$: Trajectories initiated with appropriately scaled random turbulent snapshots quickly approach a chaotic edge state, as can be seen from the temporal variation of the perturbation energy signals $\Epertrms$ depicted in \autoref{fig:E3drms_omxTri_tauWall_nosym}\textit{a,b}.
A recurrent pattern can nonetheless be found in all analysed chaotic edge trajectories:
Throughout, the dynamics are characterised by long-lasting, relatively quiescent phases of $\mathcal{O}(100\tbulk)$ length, which are suddenly interrupted by intense bursting events.
These latter manifest themselves in form of a rapid rise of the systems' perturbation energy, followed by a decrease back to the energy level of the quiescent period.
In wall-bounded turbulence, such bursting events are usually attributed to the bending and eventual breakup of a streamwise velocity streak due to the action of flanking quasi-streamwise vortices
\citep{Waleffe_1997,Hamilton_Kim_Waleffe_1995}. In order to deduce near which of the four walls such a bursting event is happening, the enstrophy contained in the streamwise-averaged vorticity \citep{Uhlmann_al_2007}
\begin{equation}
    S_i(t)=\displaystyle\int\limits_{\dom_i} \xavg{\omx}^2 \, \mathrm{d}A \qquad \quad\forall i\in\{W,S,E,N\}
  \label{eq:def_Si}
\end{equation}
is measured for the four disjoint triangular subsets 
\begin{equation}
  \begin{array}{lclc}
  \dom_{S}&=\{(y,z)|\;y<z \;\wedge\; y<-z\},&
  \dom_{N}&=\{(y,z)|\;y>z \;\wedge\; y>-z\},\\[0.5ex]
  \dom_{W}&=\{(y,z)|\;y>z \;\wedge\; y<-z\},&
  \dom_{E}&=\{(y,z)|\;y<z \;\wedge\; y>-z\}
  \end{array}
  \label{eq:dom_triangles}
\end{equation}
of the duct cross-section $\domperp=[-\hf,\hf]\times[-\hf,\hf]$ separately.
In order to quantify which of the two pairs of parallel walls features a significant turbulent activity, we define in addition a dimensionless indicator function \citep{Uhlmann_al_2007}
\begin{equation}
  \I\equiv\dfrac{(S_N + S_S) - (S_W + S_E)}{S_N+S_S+S_W+S_E},
  \label{eq:Id}
\end{equation}
based on the triangular sector contributions defined in equation~\eqref{eq:def_Si}.
By definition, $\I$ attains values close to plus or minus unity whenever the enstrophy carried by the streamwise-averaged vorticity is predominantly focussed near the walls at $y\in\{-\hf,\hf\}$ or those at $z\in\{-\hf,\hf\}$, respectively.

The temporal variation of the different contributions $S_i(t)$ presented in
\autoref{fig:E3drms_omxTri_tauWall_nosym}\textit{c,d} clearly indicate that over a significant portion of the observation interval, the turbulent vortical activity is primarily concentrated near a single wall. As has been shown earlier, this (instantaneous) `localisation' of streaks and vortices to few walls only is not exclusive to edge trajectories, but is visible in marginally turbulent flows as well
(cf. \autoref{fig:uvw_xavg_snaps_closedDuctref}\textit{a}).
Interestingly, the concentration to a single wall is strongest at the peaks of the bursting phases, with over $99\%$ (\casenosymshort{2000}) and $97\%$ (\casenosymlong{2000}) of the enstrophy associated with $\xavg{\omx}$ residing in the vicinity of a single wall during these times.
Also, \autoref{fig:E3drms_omxTri_tauWall_nosym}\textit{c,d} reveals that each bursting episode is accompanied by a `switching' of the activity to one of the neighbouring walls. During this switching period, in turn, turbulent activity spreads temporarily over several triangular sectors $\dom_{i}$, before focussing again in a single one.
The sequence in which the individual contributions $S_{i}$ dominate the overall vortical activity moreover implies that the `switching' is not necessarily continuously oriented in one direction, but that a back-and-forth switching is equally possible. %\MSw{Add `out of the scope'?}
This is inasmuch interesting as qualitatively similar chaotic dynamics is reported for the lateral `streak switching' in the chaotic edge states of the asymptotic boundary layer \citep{Khapko_al_2013}, in this case in form of an irregular left-right switching.
In the square duct case, the `wall switching' dynamics also leaves its footprint in the temporal evolution of the wall shear stress, separately averaged over the four surrounding walls:
As can be seen from \autoref{fig:E3drms_omxTri_tauWall_nosym}\textit{e,f}, the wall shear stress $\tauwi$ of `active walls' can exceed the values in a corresponding laminar flow by about $70\%$. The vicinity of the remaining walls is then usually quiescent, with an averaged wall shear stress $\tauwi\approx\tauwilam$, underlining the absence of a self-sustaining near-wall regeneration cycle along these walls.

%==============================================================
% insert figure:
\begin{figure}%[tp]
    \centering
    \includegraphics[width=0.95\linewidth]
    {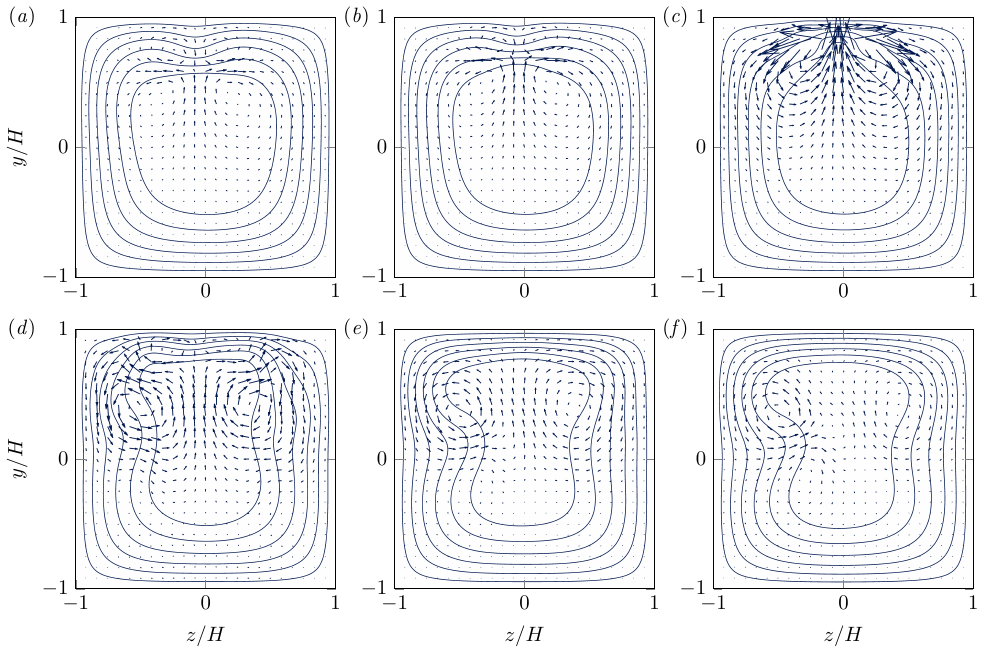}
    \caption
    {
    Selected snapshots of the instantaneous streamwise-averaged primary and secondary velocity field for case~\casenosymshort{2000}.
    The snapshots are extracted at different times along a selected bursting event $t/\tbulk\in[600,900]$,
    indicated by the markers in \autoref{fig:E3drms_omxTri_tauWall_nosym}\textit{a}.
    Isovalues are the same as in \autoref{fig:uvw_xavg_snaps_closedDuctref}.
    }
\label{fig:uvw_xavg_snaps_nosym}
\end{figure}

\begin{figure}%[tp]
    \centering
    \includegraphics[width=0.95\linewidth]
    {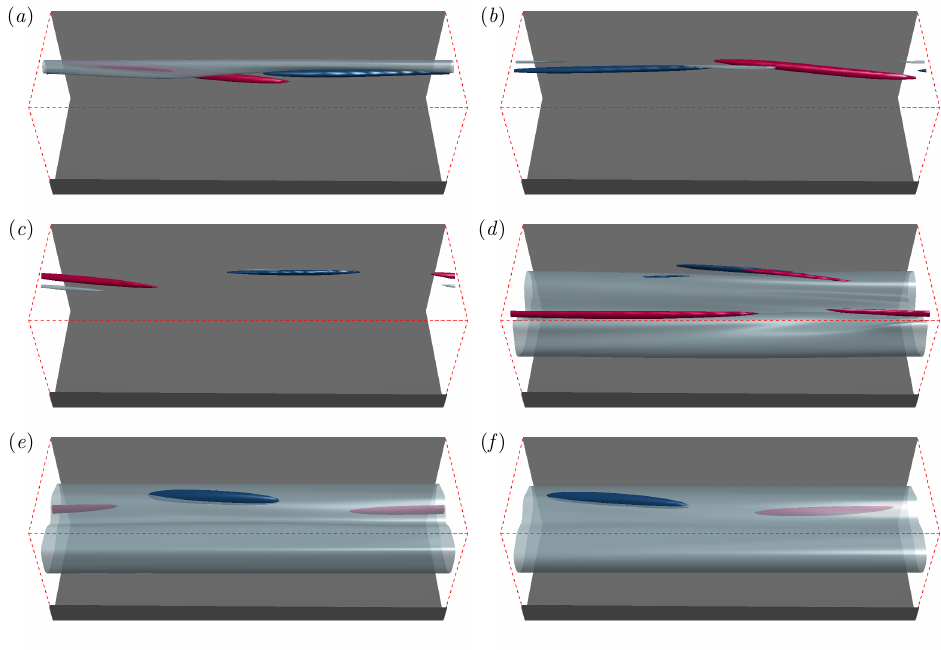}
    \caption
    {
    Three-dimensional visualisations of the velocity field in case~\casenosymshort{2000} in terms of low-speed streaks and quasi-streamwise vortices during the selected bursting event $t/\tbulk\in[600,900]$, extracted at the same times as those in \autoref{fig:uvw_xavg_snaps_nosym} and as indicated by the markers in \autoref{fig:E3drms_omxTri_tauWall_nosym}\textit{a}.
    Light blue iso-surfaces of the streamwise perturbation velocity $\upl=-0.15\ubulk$ indicate the position of low-speed streaks. Quasi-streamwise vortices are identified in terms of the $Q$-criterion of \citet{Hunt_al_1988} as regions with $Q > 0.6 \max_\dom(Q)$. Clockwise (red) and counter-clockwise rotation (dark blue) is measured by the local sign of $\omx$.
    }
\label{fig:uvw_3dview_snaps_nosym}
\end{figure}

%==============================================================
%
The different stages of a selected bursting event in the interval $t/\tbulk\in[600,900]$ and the corresponding wall switching are shown in \autoref{fig:uvw_xavg_snaps_nosym} in terms of the streamwise-averaged velocity field $\uvecxt$, extracted from the edge trajectory at the instances indicated by solid markers in
\autoref{fig:E3drms_omxTri_tauWall_nosym}\textit{a}.
\autoref{fig:uvw_3dview_snaps_nosym} complements this view with three-dimensional visualisations of the low-speed streaks (depicted as iso-surfaces of $\upl$) and the flanking quasi-streamwise vortices (measured as regions of high values of the second invariant of the velocity gradient tensor, $Q$, \citealp{Hunt_al_1988}) at the same moment in time.
\autoref{fig:uvw_3dview_snaps_nosym}\textit{a} reveals that the flow state just before the energetic burst excursion is characterised by a single pronounced low-velocity streak near the centre of the top wall, flanked by a pair of counter-rotating quasi-streamwise vortices in a streamwise staggered arrangement. The vortices are inclined w.r.t. both the neighbouring wall and the streamwise direction, so that they induce a four-vortex cell pattern in the streamwise averaged field (cf. \autoref{fig:uvw_xavg_snaps_nosym}\textit{a}). While this configuration promotes the evolution of the low-speed streak in the direct vicinity of the wall, it counteracts it further away -- until the low-speed streak eventually turns into a high-speed streak as the burst reaches its peak
(\autoref{fig:uvw_xavg_snaps_nosym}\textit{c} and \autoref{fig:uvw_3dview_snaps_nosym}\textit{c}).
Here, the special shape of the duct cross-section comes into play, since the quasi-streamwise vortices responsible for the birth of the high-speed streak at the top wall simultaneously induce new low-speed streaks along the two sidewalls at $z=\{-\hf,\hf\}$, while the energy quickly tends back to the pre-burst level
(\autoref{fig:uvw_xavg_snaps_nosym}\textit{d} and \autoref{fig:uvw_3dview_snaps_nosym}\textit{d}).
The flow is, however, not entirely symmetric w.r.t. to the $y$-axis and, eventually, a pronounced low-speed streak surrounded by a pair of quasi-streamwise vortices is only evolving on one of the sidewalls at $z=-\hf$
(\autoref{fig:uvw_xavg_snaps_nosym}\textit{e,f} and \autoref{fig:uvw_3dview_snaps_nosym}\textit{e,f}).
In contrast to their counterparts at the top wall just before the bursting event, these structures are still in an early stage of their lifetime: The streak is still aligned with the streamwise direction, while the vortices do not lean over the streak yet, as can be seen from the streamwise-averaged secondary flow patterns in \autoref{fig:uvw_xavg_snaps_nosym}\textit{e,f}.

%==============================================================
% insert figure:
\begin{figure}%[tp]
    \centering
    \includegraphics[width=0.88\linewidth]
    {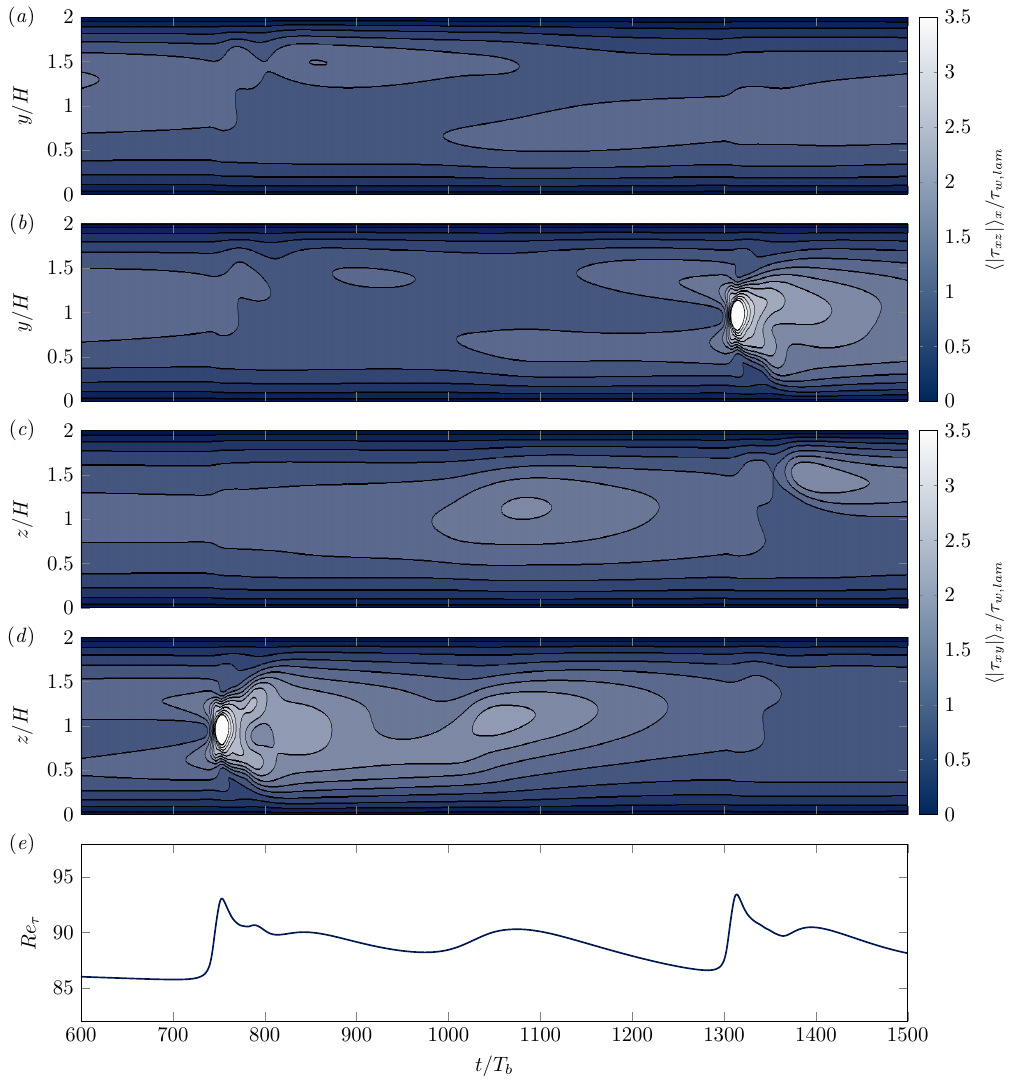}
    \caption
    {
    Variation of the streamwise-averaged wall shear stress along each wall during two bursting episodes in the interval $t/\tbulk\in[600,1500]$ of case~\casenosymshort{2000}:
    $\xavg{\tauwxz}$ along (\textit{a}) the left ($z=-\hf$) and (\textit{b}) right sidewall ($z=\hf$), as well as
    $\xavg{\tauwxy}$ along (\textit{c}) the bottom ($y=-\hf$) and (\textit{d}) top wall ($y=\hf$).
    The wall shear stress is normalised with the perimeter-averaged value in the corresponding laminar state.
    Isocontours are drawn in the interval $[0,3.5]$, with increment $0.25$.
    (\textit{e}) Temporal variation of the perimeter-averaged friction Reynolds number $\Retau$ over the same time window.
    }
\label{fig:tauw_vs_timeper_nosym}
\end{figure}

%==============================================================
%
The successive change of a mild low-speed streak to an intense high-speed region in two different bursting events
at $t\approx750\tbulk$ (top wall $y=\hf$) and $t\approx1320\tbulk$ (right sidewall $z=\hf$) can be clearly identified
by the footprint these events leave in the wall shear stress profiles in \autoref{fig:tauw_vs_timeper_nosym}.
Therein, the temporal evolution of the streamwise-averaged wall shear stress ($\abs{\tauwxy}$ or $\abs{\tauwxz}$) is shown for all four walls separately, and is contrasted with the time signal of the perimeter-averaged friction Reynolds number. The visualisation underlines that the high shear stress declines quickly in $\mathcal{O}(10\tbulk)$ after the bursting event, but its value at the corresponding wall stays the dominant contribution over a much longer time interval.
In between the two bursting events, when the turbulent activity is rather evenly distributed across the different triangular sectors, there is also no pronounced high- or low-speed streak detectable on either of the four walls.
The edge state therefore features two characteristic time scales, a slow one with $\mathcal{O}(100)\tbulk$ that is associated with the interval between two bursting events and a much faster related to the structures' downstream propagation
\citep{Kreilos_Veble_Schneider_Eckhardt_2013}.

%==============================================================
% insert figure:
\begin{figure}%[tp]
    \centering
    \includegraphics[width=0.88\linewidth]
    {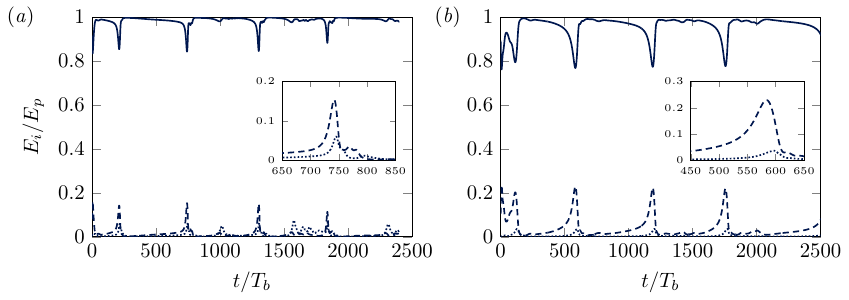}
    \caption
    {
    Time evolution of the individual contributions to the perturbation energy $\Ep$ for cases
    (\textit{a}) \casenosymshort{2000} and
    (\textit{b}) \casenosymlong{2000}.
    Line styles indicate:
    $\Estreak$ (solid),
    $\Ewave$ (dashed),
    $150\Erolls$ (dotted).
    The insets show the evolution of the wave and roll modes' amplitude enlarged during selected bursting events.
    }
\label{fig:Estreak_rolls_wave_nosym}
\end{figure}

%
%==============================================================

The observed streak-vortex interaction, including the breakup and bursting of the streaks, agrees well with the different phases of the classical self-sustained process/vortex-wave interaction envisaged by \citet{Hamilton_Kim_Waleffe_1995}, \citet{Waleffe_1997} and \citet{Hall_Sherwin_2010}. To provide quantitative evidence, the kinetic perturbation energy per unit mass is decomposed into different contributions that can be associated with the different ingredients of the process
\citep{Lucas_Kerswell_2017}, \viz
\begin{equation}
  \Ep = \dfrac{1}{2} \volavg{\uplvec^2}
      = {\underbrace{\dfrac{1}{2}\volavg{\xavg{\upl}^2}}_\text{\Estreak}}
      + {\underbrace{\dfrac{1}{2}\volavg{\xavg{\vpl}^2+\xavg{\wpl}^2}}_\text{\Erolls}}
      + {\underbrace{\dfrac{1}{2}\volavg{\left(\uplvec-\xavg{\uplvec}\right)^2}}_\text{\Ewave}}.
  \label{eq:Epert_decomp}
\end{equation}
Here, $\Estreak$, $\Erolls$ and $\Ewave$ refer to the energy contributions of the streamwise constant streak mode, the quasi-streamwise vortices/rolls and the streamwise-varying wave mode, respectively. In the square duct, $\Erolls$ effectively measures the kinetic energy of the instantaneous secondary flow.
Despite being the dominant contribution throughout, \autoref{fig:Estreak_rolls_wave_nosym} highlights that the intensity of the streamwise-elongated streaks $\Estreak$ regularly drops during the identified bursting intervals.
In these phases, the energy is instead transferred into the streamwise-dependent wave field ($\Ewave$), that is, the initially streamwise-aligned streak gets more and more distorted. With a slight delay of several bulk time units, a rise of the roll mode energy $\Erolls$ follows, which indicates the gain in strength of the quasi-streamwise vortices. Owing to the much smaller amplitude of the cross-stream velocity components compared to their streamwise counterpart, the kinetic energy residing in the quasi-streamwise vortices is two orders of magnitude smaller than that of the wave field. The outlined observations are consistent for the shorter and longer domain, with somewhat higher portions of the total perturbation energy $\Ep$ being held in $\Ewave$ for the longer domain.

%==============================================================
The fact that the edge state in the square duct is a chaotic attractor within $\edgemfd$ is consistent with circular pipe flows, where the edge state takes the same form
\citep{Schneider_Eckhardt_2009,Duguet_Willis_Kerswell_2008,Avila_Barkley_Hof_2023,%
       Mellibovsky_Meseguer_Schneider_Eckhardt_2009}.
Given that many TW solutions detected in circular pipe flow \citep{Pringle_Duguet_Kerswell_2009} possess a strikingly similar counterpart in the square duct \citep{Okino_2011}, the current findings provide further support for the suspected close relation between the state space structure of both systems.
When it comes to a direct comparison of the individual dynamics, however, it turns out that comparable alternating quiescent phases and intermittent bursts have not been reported for the chaotic edge states of circular pipe flows.
In other systems including plane channels
\citep{Toh_Itano_2003,Rawat_Cossu_Rincon_2014,Rawat_Cossu_Rincon_2016,Zammert_Eckhardt_2014}
or asymptotic suction boundary layers \citep{Kreilos_Veble_Schneider_Eckhardt_2013}, though, the edge states do feature a remarkably similar dynamics. For certain (streamwise and spanwise) domain periods and Reynolds numbers, the edge state in these systems even simplifies to a symmetric time-periodic limit cycle within $\edgemfd$, or an unstable periodic orbit w.r.t. the full state space; a more detailed discussion of these solutions and comparison with the edge states in square duct will follow in \S~\ref{subsec:sym_edge}.

%======================================================================================================================%
%----------------------------------------------------------------------------------------------------------------------%
%======================================================================================================================%

\subsection{Periodic edge-state dynamics in symmetric subspaces}\label{subsec:sym_edge}
In circular pipe flow, edge states reduce to simple invariant solutions merely if trajectories are restricted to appropriate symmetric subspaces of the full state space. Examples include travelling wave edge states in $\pi$-rotational symmetric flows \citep{Duguet_Willis_Kerswell_2008} and (relative) periodic orbit edge states under enforcement of a $\pi$-rotational, combined with a reflectional symmetry \citep{Avila_al_2013}.
Based on these experiences, \citet{Gepner_Okino_Kawahara_2025} (cf. also \citealp{Okino_2014}, in Japanese) performed edge-tracking studies in the $\pirot$-symmetric subspace of square duct flows driven with a constant pressure gradient.
For domains of $\Lx/\hf=2\pi$ and $\Lx/\hf=8\pi$ length, these authors found the edge states to be travelling waves equivariant under the symmetry groups $(\yshrefl,\zshrefl,\pirot)$ and $(\yrefl,\zrefl,\pirot)$, respectively.
For the intermediate domain length $\Lx/\hf=4\pi$, though, the edge state was chaotic even in the $\pirot$-symmetric subspace.
In the current study, we found a travelling wave edge state with the same characteristics as the one found by \citet{Okino_2014} in the $\pirot$-symmetric subspace despite our somewhat different driving. Both solutions arguably belong to the same family of solutions, proving the consistency of our results with theirs.

In the remainder of this section, we restrict ourselves to the $\yrefl$- and $\yrefl\zrefl$-symmetric subspaces.
Under these symmetry constraints, edge trajectories at various Reynolds numbers in the range $\Reb\in[1600,3200]$ converge to limit cycles. At selected Reynolds numbers, edge trajectories have been computed starting from different turbulent snapshots as initial conditions.
For almost all parameter points, the different edge trajectories eventually converge to the same periodic edge states modulo discrete or continuous symmetry operations.
The only exception is the $\yrefl$-symmetric case~\caseyrefl{2000}, for which one edge trajectory converged to the periodic orbit shown in \autoref{fig:E3drms_omxTri_yrefl}\textit{b}, while the other approached a travelling wave. The latter has been successfully converged with a standard Newton-Raphson method, using the generalised minimal residual (GMRES) approach to approximately solve the arising linear systems \citep{Chandler_Kerswell_2013}.
The detected travelling wave lives in a higher-symmetric subspace invariant under $(\yrefl,\zshrefl,\pishrot)$ and presumably belongs to the family of solutions found by \citet{Wedin_Bottaro_Nagata_2009}, although no effort was made to prove this rigorously via a continuation study.
To show that the travelling wave is indeed an alternative edge state rather than a relative saddle within $\edgemfd$, we continued tracking the edge trajectory for several thousand bulk time units after it had approached the travelling wave, without any sign of deviation from the latter. 
State space configurations with multiple co-existing `local' edge states of possibly varying type (e.g. one travelling wave and one periodic orbit as observed here) are not exclusive to the square duct, but have been reported for a variety of flows \citep{Duguet_Willis_Kerswell_2008,Suri_Pallantla_Schatz_Grigoriev_2019}.

In the remainder, we will restrict our analysis to the periodic edge states.
Unfortunately, the limit cycles' long periods $T=\mathcal{O}(100\tbulk)$ do not allow to converge
these solutions to machine precision with the standard Newton-GMRES approach; a problem that is shared with time-periodic edge states in other systems \citep{Kreilos_Veble_Schneider_Eckhardt_2013,Zammert_Eckhardt_2014}.
If at all possible, convergence might be achieved with more advanced multipoint-shooting techniques \citep{Budanur_al_2017}, or recently developed adjoint-based variational Jacobian-free methods \citep{Ashtari_Schneider_2023}.
Here, however, time-stepping alone provides a fairly good approximation to the periodic orbit since the latter is an attracting state in $\edgemfd$:
The relative distance between two velocity fields separated by a full period $T$, \viz
\begin{equation}
 \min_{\xshift}\dfrac{\Lnorm{\uvec(\xvec+\xvecshift,t+T)-\uvec(\xvec,t)}{2}}{\Lnorm{\uvec(\xvec,t)}{2}}
 \quad \text{with} \; \xvecshift=(\xshift,0,0)^T,
 \label{eq:L2dist_tper}
\end{equation}
is small -- of order $\mathcal{O}(10^{-5})$ when only the streamwise-dependent modes $\uvec-\xavg{\uvec}$ are compared in equation~\eqref{eq:L2dist_tper}, and $\mathcal{O}(10^{-8})$ if the full set of modes is considered.
Here, the period of the limit cycles is estimated up to a precision of $5\deltat \approx 0.07\tbulk$ (which is half the interval in which statistics have been aggregated) based on the peaks of the auto-correlation functions of different measures; see \autoref{app:appendix_conv} for further details and \autoref{app:appendix_stab} for an explanation how the stability of the limit cycles has been verified based on first-return maps of $\I(t)$ \citep{Khapko_al_2013,Zammert_Eckhardt_2014}.
Note that the minimisation over all possible streamwise shifts $\xshift$ in equation~\eqref{eq:L2dist_tper} checks for relative periodic orbits, which are self-recurrent in a co-moving frame of reference only. It turns out that the identified edge states are relative periodic orbits indeed:
In case~\caseyrefl{2400}, for instance, $\uvec(\xvec+\xvecshift,t+T)=\uvec(\xvec,t)$ holds for $\xshift/\hf\approx0.11$.

\subsubsection{\edgemfd{} under $\yrefl$-symmetry}\label{subsec:yrefl_edge}
%
%==============================================================
% insert figure:
\begin{figure}%[tp]
    \centering
    \includegraphics[width=0.88\linewidth]
    {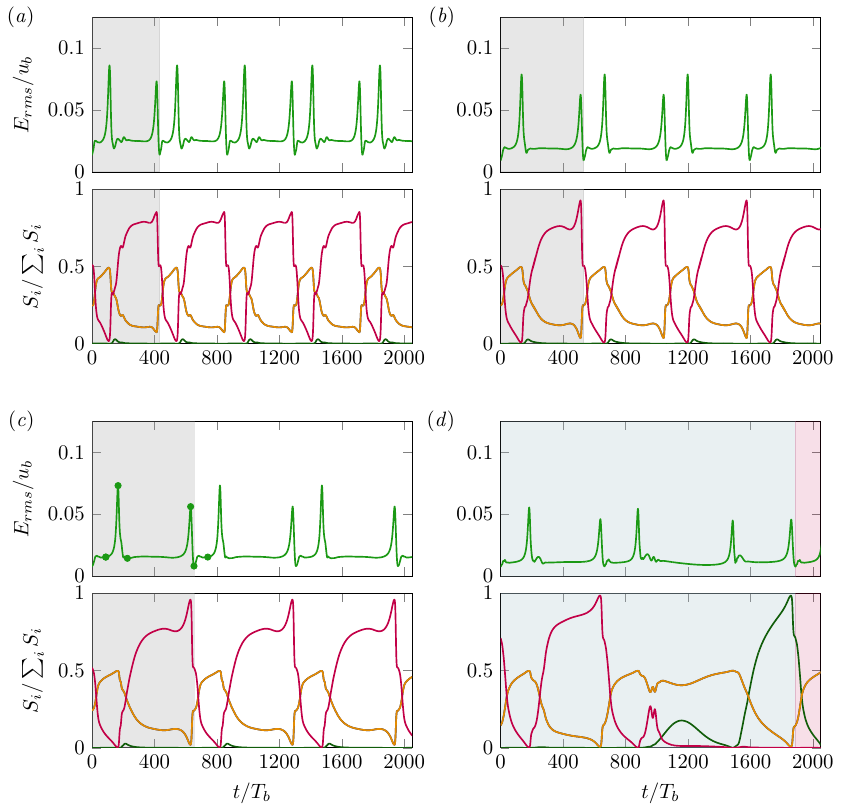}
    \caption
    {
    Time evolution of the r.m.s. energy signal $\Epertrms(t)$ (upper plots) and individual contributions of the triangular sectors to the overall mean streamwise enstrophy $S_i(t)$ (lower plots) for periodic edge states in the $\yrefl$-symmetric subspace:
    (\textit{a}) \caseyrefl{1600},
    (\textit{b}) \caseyrefl{2000},
    (\textit{c}) \caseyrefl{2400} and
    (\textit{d}) \caseyrefl{3200}.
    In (\textit{a-c}), the fundamental period (grey background) is repeated periodically.
    For the sake of visualisation, in (\textit{d}), we show only the first half of the POs period (blue background)
    which represents a per-periodic orbit with period $T_{pre}=\tper/2$ (see the discussion in the main text).
    Colour coding in the lower plots is:
    $i={E}$ (green),
    $i={W}$ (red),
    $i=\{S,N\}$ (orange, identical due to symmetry).
    }
\label{fig:E3drms_omxTri_yrefl}
\end{figure}

%==============================================================
%==============================================================
% insert figure:
\begin{figure}%[tp]
    \centering
    \includegraphics[width=0.95\linewidth]
    {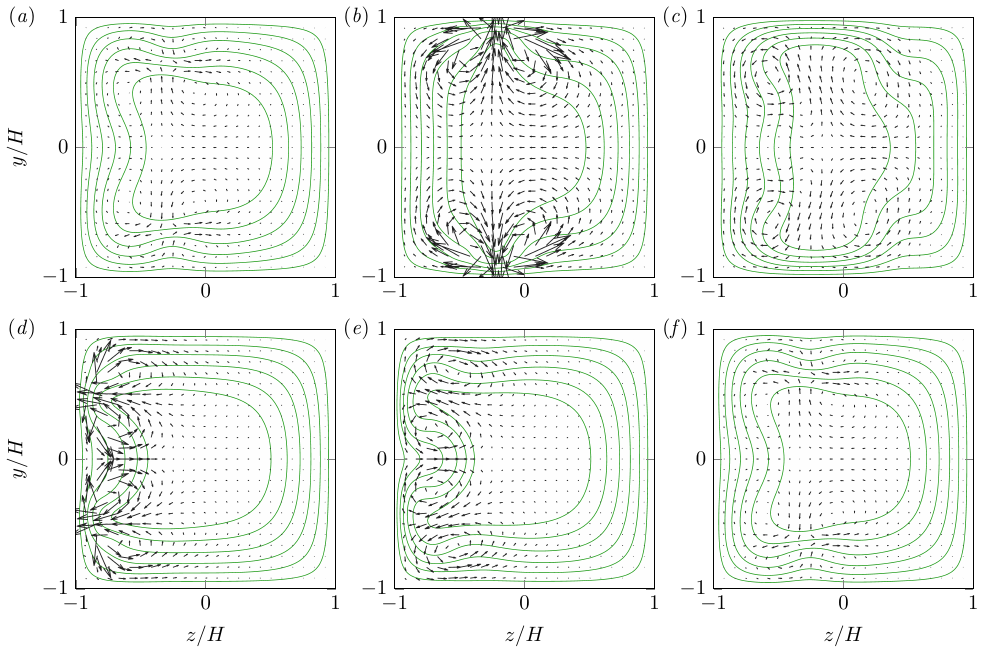}
    \caption
    {
    Snapshots of the streamwise-averaged primary and secondary flow field of case~\caseyrefl{2400} extracted at different instances along the orbit, as
    indicated by the markers in \autoref{fig:E3drms_omxTri_yrefl}\textit{c}.
    Isovalues are the same as in \autoref{fig:uvw_xavg_snaps_closedDuctref}.
    }
\label{fig:uvw_xavg_snaps_yrefl}
\end{figure}

%==============================================================
%==============================================================
% insert figure:
\begin{figure}%[tp]
    \centering
    \includegraphics[width=0.88\linewidth]
    {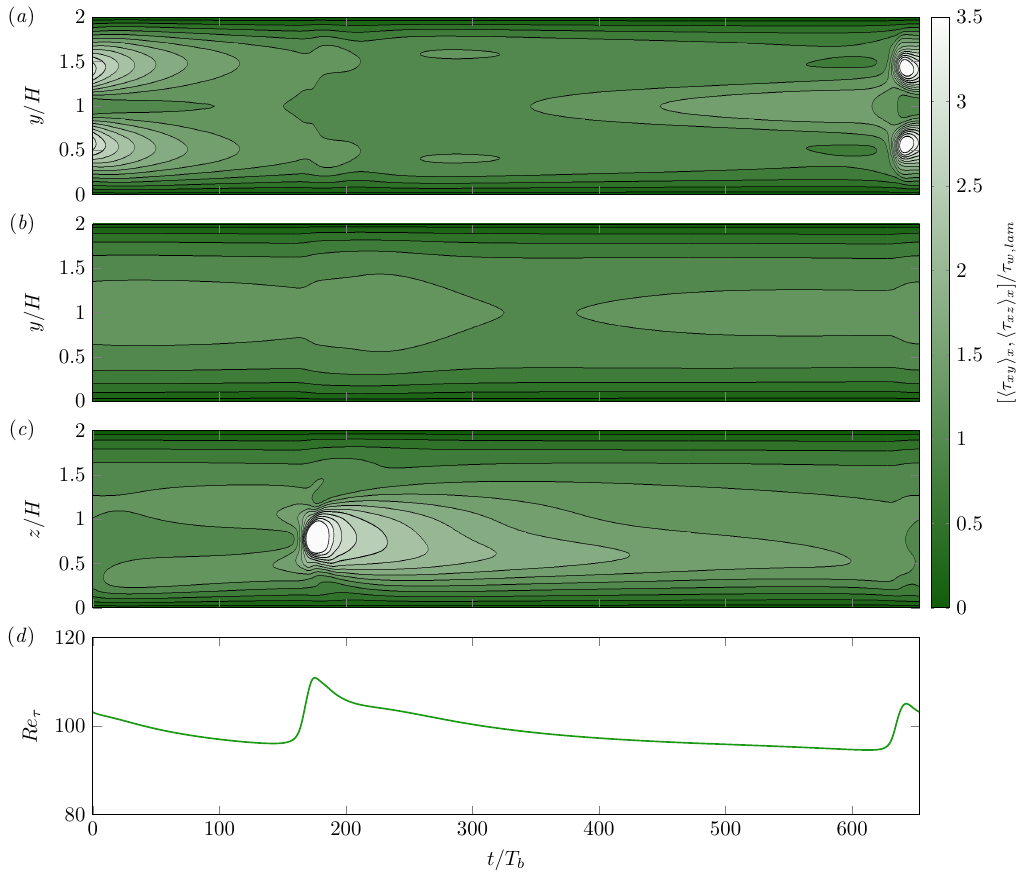}
    \caption
    {
    Variation of the streamwise-averaged wall shear stress along each wall over a full period of case~\caseyrefl{2400}:
    $\xavg{\tauwxz}$ along (\textit{a}) the left sidewall ($z=-\hf$) and (\textit{b}) right sidewall ($z=\hf$) and (\textit{c}) $\xavg{\tauwxy}$ along the top/bottom wall ($y=\pm\hf$).
    The wall shear stress is normalised with the perimeter-averaged value in the corresponding laminar state.
    Isocontours are drawn in the interval $[0,3.5]$, with increment $0.25$.
    (\textit{d}) Temporal variation of the perimeter-averaged friction Reynolds number $\Retau$ over the same time window.
    }
\label{fig:tauw_vs_timeper_yrefl}
\end{figure}

%==============================================================
%
%
In a first step, we analyse states whose velocity and pressure fields obey a single reflectional symmetry w.r.t. the wall bisector. As for the chaotic edge states shown in \autoref{fig:E3drms_omxTri_tauWall_nosym}\textit{a,b}, the edge trajectories in the $\yrefl$-symmetric subspace experience regular bursting events that interrupt the otherwise quiescent dynamics (cf. \autoref{fig:E3drms_omxTri_yrefl}, upper plots in each panel) for all considered Reynolds numbers.
Per periodic cycle, the flow undergoes -- depending on the Reynolds number -- between two and ten such bursting episodes with different peak amplitudes. In all cases, the strongest bursts go hand in hand with a rapid increase in energy by approximately a factor three compared to the pre-burst phase.

The temporal variation of the individual sector contributions to the streamwise-averaged enstrophy, $S_i(t)$, is presented in \autoref{fig:E3drms_omxTri_yrefl} (lower plots in each panel). For all but case~\caseyrefl{3200}, the two consecutive peaks of the energy signals come with a subsequent switching from a `top-bottom' (higher peak) to a `left-only' (weaker peak) configuration, or vice versa. The two configurations are seen in the streamwise-averaged velocity fields, extracted along a full cycle of case~\caseyrefl{2400} (cf. \autoref{fig:uvw_xavg_snaps_yrefl}):
The higher energetic peak encodes the simultaneous bursting of a streak at the top wall and its symmetric twin at the bottom wall (\autoref{fig:uvw_xavg_snaps_yrefl}\textit{b}), while the lower indicates a bursting event at one of the sidewalls (\autoref{fig:uvw_xavg_snaps_yrefl}\textit{d}). The streak-vortex groups at the top and bottom walls arguably undergo the same self-sustaining process as in the chaotic edge state, with the difference that the symmetrical constrains here enforce them to appear in pairs at the two opposite walls.
In analogy to the dynamics of the chaotic edge state, each of the quasi-streamwise vortices responsible for the change of the low-speed into a high-speed streak at the bottom and top wall induces a low-speed streak at the neighbouring sidewall -- a first one in the lower half ($y/\hf<0$) and a second one in the upper half ($y/\hf>0$) of the domain (\autoref{fig:uvw_xavg_snaps_yrefl}\textit{c}).
The two mirror-symmetric streaks then experience a similar instability as their larger counterparts at the top and bottom wall, until they finally break (\autoref{fig:uvw_xavg_snaps_yrefl}\textit{d}) and the associated quasi-streamwise vortices give rise to a new generation of streaks at the top and bottom walls (\autoref{fig:uvw_xavg_snaps_yrefl}\textit{f}).
A comparison of this last state in \autoref{fig:uvw_xavg_snaps_yrefl}\textit{f} with the one in \autoref{fig:uvw_xavg_snaps_yrefl}\textit{a} once more confirms the exact recurrence of the orbit after a period $\tper=653.3\tbulk$.

The switching behaviour between a single streak-vortex family at the bottom and top wall on the one hand and the two narrow streaks with associated vortices at one of the sidewalls leaves a footprint in the temporal evolution of the streamwise-averaged wall shear stress, cf. \autoref{fig:tauw_vs_timeper_yrefl}. On the other hand, the presented data also indicates that the remaining wall at $z=\hf$ never hosts a pronounced streak, again highlighting that the switching occurs only between the top/bottom walls and the sidewall at $z=-\hf$ in this case.
The observed behaviour is qualitatively similar for the lower Reynolds number cases~\caseyrefl{1600}, \caseyrefl{2000}, \caseyrefl{2200}(not shown) and \caseyrefl{2400}, the only difference being that the period of the orbits monotonically increases with the bulk Reynolds number (cf. discussion of \autoref{fig:Tper_vs_Reb}). Such a trend is consistent with observations for periodic edge states in other flow configurations \citep{Kreilos_Veble_Schneider_Eckhardt_2013}.
These characteristics change for the highest considered Reynolds number (case~\caseyrefl{3200}), for which the edge state is still periodic, but the dynamics are more complex and the period is with about $3800\tbulk$ one order of magnitude longer than those at the lower Reynolds numbers.
In contrast to these latter cases, the edge state in case~\caseyrefl{3200} switches between three rather than two distinct states in a fairly complicated manner (cf. \autoref{fig:E3drms_omxTri_yrefl}\textit{d}). The sequence of dominant contributions $S_i$ over a single cycle reads
\begin{equation*}
    \underbrace{(S_S\cup S_N) \rightarrow S_W \rightarrow (S_{S}\cup S_{N}) \rightarrow S_W}_{\text{pre-PO}_1}
    \rightarrow
    \underbrace{(S_S\cup S_N) \rightarrow S_E \rightarrow (S_{S}\cup S_{N}) \rightarrow S_E}_{\text{pre-PO}_2},
\end{equation*}
i.e. the flow changes between a `top-bottom' state and a flow state with turbulent vortical activity focussing near either of the two sidewalls. It turns out that the two sequences marked as $\text{pre-PO}_1$ and $\text{pre-PO}_2$
represent two pre-periodic orbits \citep[][pp.~187]{Cvitanovic_chaosbook}: After $\tperpre=\tper/2$, the flow state repeats itself modulo a reflection such that $\uvec(\xvec,t) = \zrefl\uvec(\xvec,t + \tperpre)$ and
$\uvec(\xvec,t) = \zrefl\zrefl\uvec(\xvec,t + \tper) = \uvec(\xvec,t + \tper)$ holds;
ignoring possible streamwise shifts for readability.
Note that such sudden changes of the edge state characteristics due to parameter variations (e.g. changing domain extents or Reynolds number) are reported for edge states in many flow configurations, sometimes even changing from a simple travelling wave or periodic orbit to a chaotic state \citep{Zammert_Eckhardt_2014,Kreilos_Veble_Schneider_Eckhardt_2013,Khapko_al_2013}.

\subsubsection{\edgemfd{} under $\yrefl\zrefl$-symmetry}\label{subsec:yzrefl_edge}
%==============================================================
% insert figure:
\begin{figure}%[tp]
    \centering
    \includegraphics[width=0.88\linewidth]
    {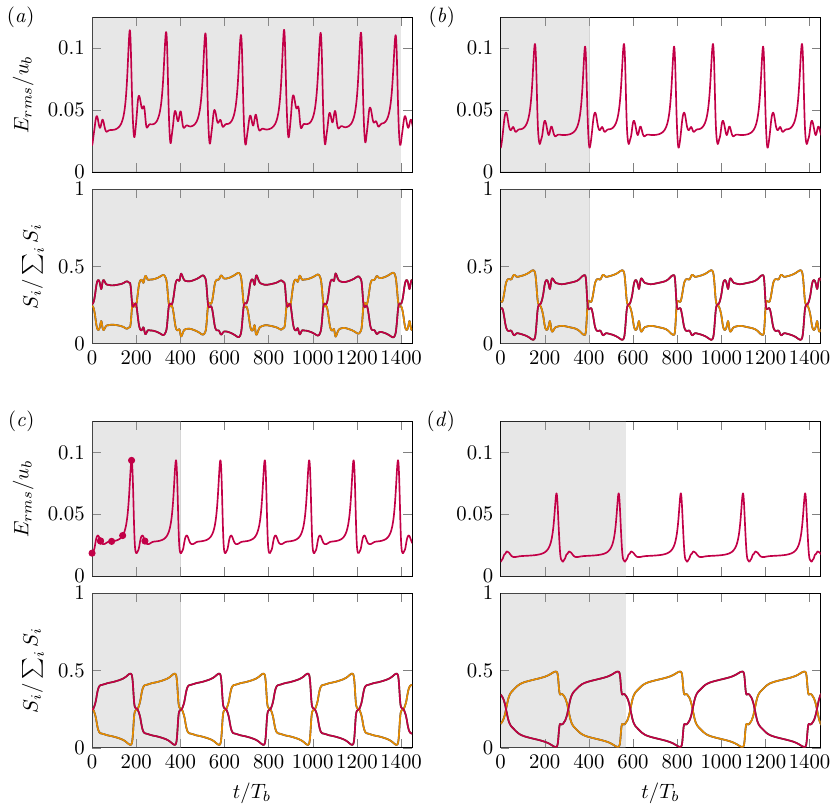}
    \caption
    {
    Time evolution of the r.m.s. energy signal $\Epertrms(t)$ (upper plots) and individual contributions of the triangular sectors to the overall mean streamwise enstrophy $S_i(t)$ (lower plots) for periodic edge states in the $\yrefl\zrefl$-symmetric subspace:
    (\textit{a}) \caseyzrefl{1600},
    (\textit{b}) \caseyzrefl{1800},
    (\textit{c}) \caseyzrefl{2000} and
    (\textit{d}) \caseyzrefl{3200}.
    Colour coding in the lower plots is:
    $i=\{W,E\}$ (red, identical due to symmetry),
    $i=\{S,N\}$ (orange, identical due to symmetry).
    }
\label{fig:E3drms_omxTri_yzrefl}
\end{figure}

%==============================================================
%==============================================================
% insert figure:
\begin{figure}%[tp]
    \centering
    \includegraphics[width=0.88\linewidth]
    {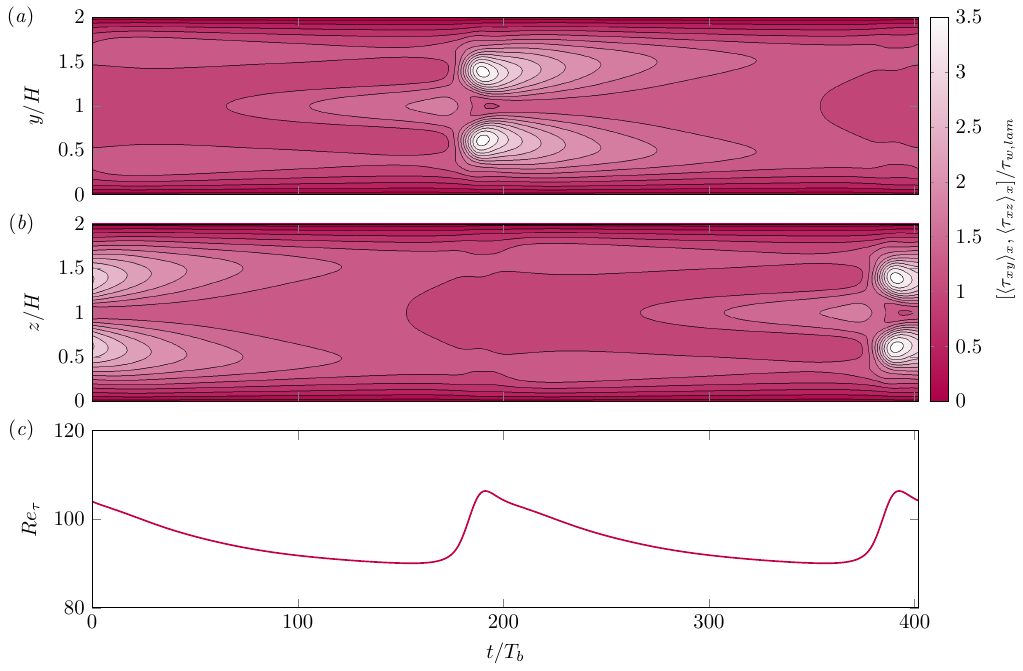}
    \caption
    {
    Variation of the streamwise-averaged wall shear stress over a full period of case~\caseyzrefl{2000}:
    (\textit{a}) $\xavg{\tauwxz}$ along the left/right sidewalls ($z=\pm\hf$) and
    (\textit{b}) $\xavg{\tauwxy}$ along the top/bottom walls ($y=\pm\hf$).
    (\textit{c}) Temporal variation of the perimeter-averaged friction Reynolds number $\Retau$ over the same time window. Contour levels and normalisation are identical to those in \autoref{fig:tauw_vs_timeper_yrefl}.
    }
\label{fig:tauw_vs_timeper_yzrefl}
\end{figure}

%==============================================================
%==============================================================
% insert figure:
\begin{figure}%[tp]
    \centering
    \includegraphics[width=0.95\linewidth]
    {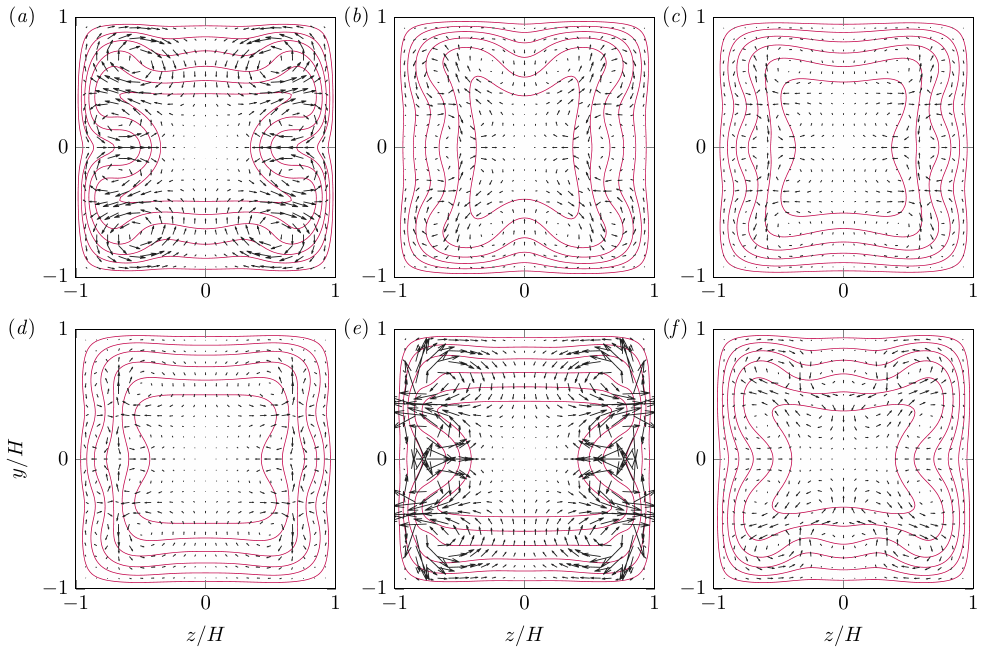}
    \caption
    {
    Snapshots of the streamwise-averaged primary and secondary flow field of case~\caseyzrefl{2000}, extracted at different instances along the orbit as
    indicated by the markers in \autoref{fig:E3drms_omxTri_yzrefl}\textit{c}.
    Isovalues are the same as in \autoref{fig:uvw_xavg_snaps_closedDuctref}.
    }
\label{fig:uvw_xavg_snaps_yzrefl}
\end{figure}

%==============================================================
%
In a second step, the flow is further restricted to the twofold mirror-symmetric subspace for which trajectories are invariant under the operations $\yrefl$ and $\zrefl$. It is then readily shown that states in this subspace automatically fulfil a rotational symmetry by $\pi$ about the $x$-axis ($\pirot$) as well.
As for the $\yrefl$-symmetric cases, all considered trajectories at various Reynolds numbers converge after an initial transient to a relative periodic orbit as edge state. The converged limit cycles can be identified in terms of $\Epertrms$ in \autoref{fig:E3drms_omxTri_yzrefl} for different Reynolds numbers. The general dynamics is similar to that of their counterparts in the $\yrefl$-symmetric subspace in that quiescent phases alternate again with intermittent intense high energy and dissipation excursions. Here, however, the systems oscillate periodically between two states in which the overall turbulent activity is almost entirely focussed at the top and bottom walls at $y=\pm\hf$ or the two sidewalls at $z=\pm\hf$, respectively.
All edge states with $\Reb\geq2000$ (cf. cases~\caseyzrefl{2000} and \caseyzrefl{3200} in \autoref{fig:E3drms_omxTri_yzrefl}\textit{c,d} as examples) undergo two bursting events with identical peak amplitudes per limit cycle, from which one is associated with a streak breakup at the top/bottom walls and the other one with a corresponding event at the sidewalls.
\autoref{fig:tauw_vs_timeper_yzrefl}, in which the temporal evolution of the streamwise-averaged wall shear stress is presented for case~\caseyzrefl{2000}, reveals that not only the peak amplitudes are identical: The limit cycle consists of two pre-periodic orbits with $\tperpre = \tper/2$ that are identical modulo $\pihalfrot$ or $\pithrhalfrot$, i.e. rotations by $\pi/2$ or $3\pi/2$ about the $x$-axis.
Hence, the relative distance between two velocity fields separated by $\tperpre$ (following the definition in equation~\eqref{eq:L2dist_tper} including a correction for the rotation) is with $\mathcal{O}(10^{-5})$ of the same order as when evaluated for the full period $T$.
In \autoref{fig:uvw_xavg_snaps_yzrefl}, therefore, instantaneous streamwise-averaged velocity fields are shown for the first half of the limit cycle only.
A comparison between the states in \autoref{fig:uvw_xavg_snaps_yzrefl}\textit{b,f} visually confirms that any two states separated by $\tperpre$ are identical except for a rotation by $\pi/2$.
Apart from that, the snapshots underline that the different phases the two neighbouring streaks undergo during a cycle of the pre-periodic orbit on each of the two sidewalls are arguably the same as those observed for the streaks along the sidewall of case~\caseyrefl{2400} in \autoref{fig:uvw_xavg_snaps_yrefl} -- enforced by the symmetry constraints to appear simultaneously on both opposing walls.

It is interesting to see that a reduction of the Reynolds number below $2000$ breaks the discussed $\pihalfrot$-symmetry between the two pre-periodic segments: In case~\caseyzrefl{1800} (cf. \autoref{fig:E3drms_omxTri_yzrefl}\textit{b}), the two energetic peaks associated with the different pairs of opposite walls differ slightly in amplitude, and the dynamics during the decline of the energy after the burst is somewhat different too.
Analogue symmetry breaks have been reported for edge states in the asymptotic boundary layer by \citet{Kreilos_Veble_Schneider_Eckhardt_2013}, in their case, though, induced by a variation of the domain size.
At even lower Reynolds numbers (cf. case~\caseyzrefl{1600}, \autoref{fig:E3drms_omxTri_yzrefl}\textit{a}), the energy signal implies that a period-doubling sequence might occur for $\Reb\lesssim1800$: A full cycle period now features eight individual bursting episodes, four associated with the `top/bottom-wall states' and four with the `sidewall states'.
%
%==============================================================
% insert figure:
\begin{figure}%[tp]
    \centering
    \includegraphics[width=0.95\linewidth]
    {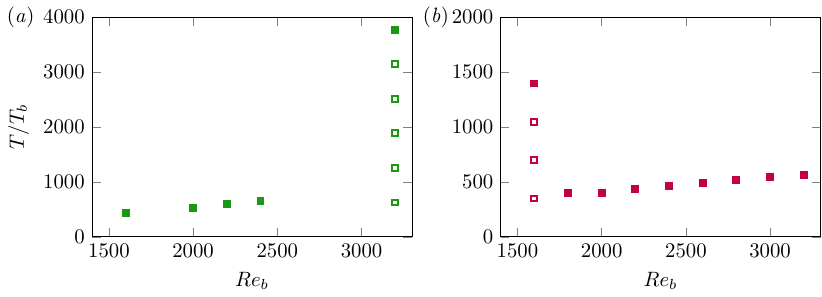}
    \caption
    {
    %\MS{Move to Appendix?}
    Edge state periods (closed squares) as a function of the bulk Reynolds number for
    (\textit{a}) the  $\yrefl$-symmetric and
    (\textit{b}) the  $\yrefl\zrefl$-symmetric subspace.
    For limit cycles with long periods, additional open symbols indicate the time separation to
    other local maxima of the auto-correlation function associated with the signal $\I(t)$.
    % \  $\tper/n$ with $n\in\natn_{>1}$.}
    %
    }
\label{fig:Tper_vs_Reb}
\end{figure}

%==============================================================
%
To completely clarify whether such a sequence takes place in the interval
$\Reb\in(1600,1800)$, many more edge-tracking runs have to be launched to resolve the `continuation' in Reynolds number -- an undertaking that is out of the scope of the current study.
Nonetheless, the results indicate that the variation of the edge state characteristics with the Reynolds number is, as for the $\yrefl$-symmetric subspace, not uniform across the parameter space: \autoref{fig:Tper_vs_Reb} visualises the edge state periods as a function of the Reynolds number, clearly implying that the period varies essentially linearly in some parameter ranges. In others, though, a similar change in Reynolds number causes abrupt increases of the limit cycle's period.

%===================================================================
  \section{Discussion}\label{sec:discussion}
  \subsection{Comparison with marginal duct turbulence}\label{sec:discuss_turb}
%
%
%==============================================================
% insert figure:
\begin{figure}%[tp]
    \centering
    \includegraphics[width=0.95\linewidth]
    {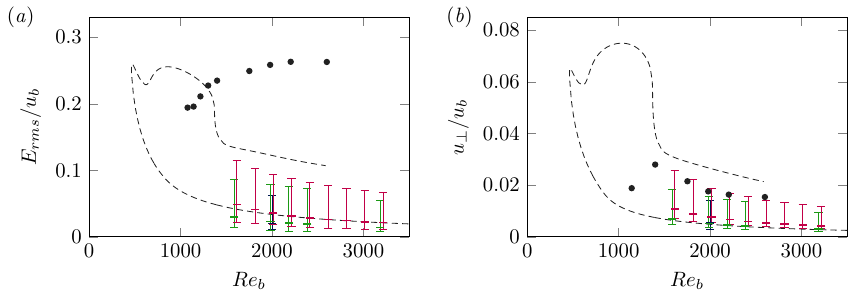}
    \caption
    {
    (\textit{a}) R.m.s. energy signal $\Epertrms/\ubulk$ and
    (\textit{b}) mean secondary flow intensity $\usec/\ubulk$
    %(\textit{a}) Mean friction Reynolds number $\Retau$,
    %(\textit{d}) period of the POs
    of the edge states as a function of the bulk Reynolds number $\Reb$.
    The vertical `error bars' visualise the range of values attained during one cycle of the periodic cases or along the
    entire chaotic edge trajectory. Time averages over a full cycle (periodic) or the full trajectory (chaotic) are indicated by a short thick horizontal line along the error bar.
    %For the two simulation series conducted for the symmetric subspaces,
    %
    %In (\textit{b,c}),
    The black dashed lines indicate the TW family found by \citet{Uhlmann_al_2010} for the same streamwise period
    $\alp\hf=1$, while black circles indicate long-time averages of turbulent trajectories
    \citep{Uhlmann_al_2007}.
    Note that data points for $\yrefl$- and $\yrefl\zrefl$-symmetric cases have been shifted to $\Reb\pm15$, respectively, to improve their visibility.
    %\MSw{ in d, add further peaks for large $T$ that indicate `almost recurrences' ...}
    %
    }
\label{fig:Retau_Erms_Evw_vs_Reb}
\end{figure}

%==============================================================
%
In \autoref{fig:Retau_Erms_Evw_vs_Reb}, we compare the perturbation energy $\Epertrms$ and the mean secondary flow intensity $\usec=\sqrt{2\Erolls}$ of the detected edge states with the corresponding values in fully-turbulent trajectories and those for the `eight-vortex' state travelling wave family of \citet{Uhlmann_al_2010} in a domain of identical length.
The key difference between the two measures is that $\Epertrms$ quantifies the energy in the streamwise-varying modes of all three velocity components, whereas $\usec$ measures the streamwise-constant modes of the cross-stream velocity components only.
From \autoref{fig:Retau_Erms_Evw_vs_Reb}, it can be seen that the period-averaged values for both quantities in the newly detected edge states are comparable to those of the TWs found by \citet{Uhlmann_al_2010}.
Instantaneously, however, the peak values of both quantities clearly exceed the corresponding period-averaged values and those attained by the travelling wave solution. While the maxima of the perturbation energy $\Epertrms$ along a periodic cycle remain clearly below the level of the turbulent long-time averages, the largest values of the mean secondary flow intensity $\usec$ are of similar amplitude as the turbulent long-time averages, in particular for the doubly-symmetric edge states. The latter is expected as the twofold symmetry enforces vortices to appear near two of the four surrounding walls during the entire cycle, while the chaotic edge state features vortical activity near a single wall most of the time, resulting in a lower mean strength of the cross-sectional velocity components.

%
%
%==============================================================
% insert figure:
\begin{figure}%[tp]
    \centering
    \includegraphics[width=0.96\linewidth]
    {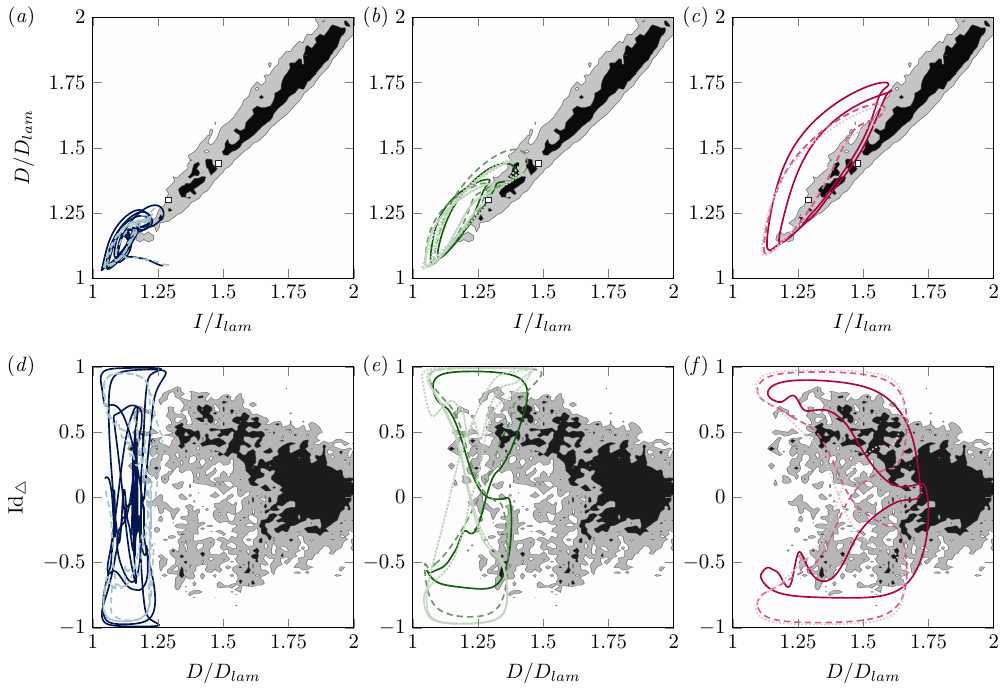}
    \caption{
    %\MSw{TBD: Longer statistics to get nicer jpdf plots?}
    State space projections onto the planes spanned by
    (\textit{a-c}) the dissipation $\Diss$ and the energy input $\Einp$ and
    (\textit{d-f}) the indicator function $\I$ and the dissipation $\Diss$.
    Energy input and dissipation are normalised with the respective values in a laminar flow with identical mass flow rate.
    The coloured curves in (\textit{a,d}) represent the dynamics along the chaotic edge states
    of cases~\casenosymshort{2000} (solid) and \casenosymlong{2000} (dashed).
    The remaining plots visualise the dynamics of the periodic edge states in
    (\textit{b,e}) the $\yrefl$-symmetric subspace (\caseyrefl{1600}, \caseyrefl{2400}, \caseyrefl{3200}
    as solid, dashed, dotted lines) and
    (\textit{c,f}) the $\yrefl\zrefl$-symmetric subspace (\caseyzrefl{1800}, \caseyzrefl{2600}, \caseyzrefl{3200}
    as solid, dashed, dotted lines).
    The gray-coloured background maps represent joint probability functions of the same quantities in the turbulent reference simulation at $\Reb=1150$, with contours enclosing $90\%$ and $50\%$ of their total masses.
    The white-filled squares in (\textit{a-c}) represent the two flow states in
    \autoref{fig:uvw_xavg_snaps_closedDuctref}\textit{a,b}, with the lower dissipation state being the `two-vortex state' in \autoref{fig:uvw_xavg_snaps_closedDuctref}\textit{a}.
    }
\label{fig:Id_Einp_vs_Diss}
\end{figure}

%==============================================================
\begin{figure}%[tp]
    \centering
    \includegraphics[width=0.95\linewidth]
    {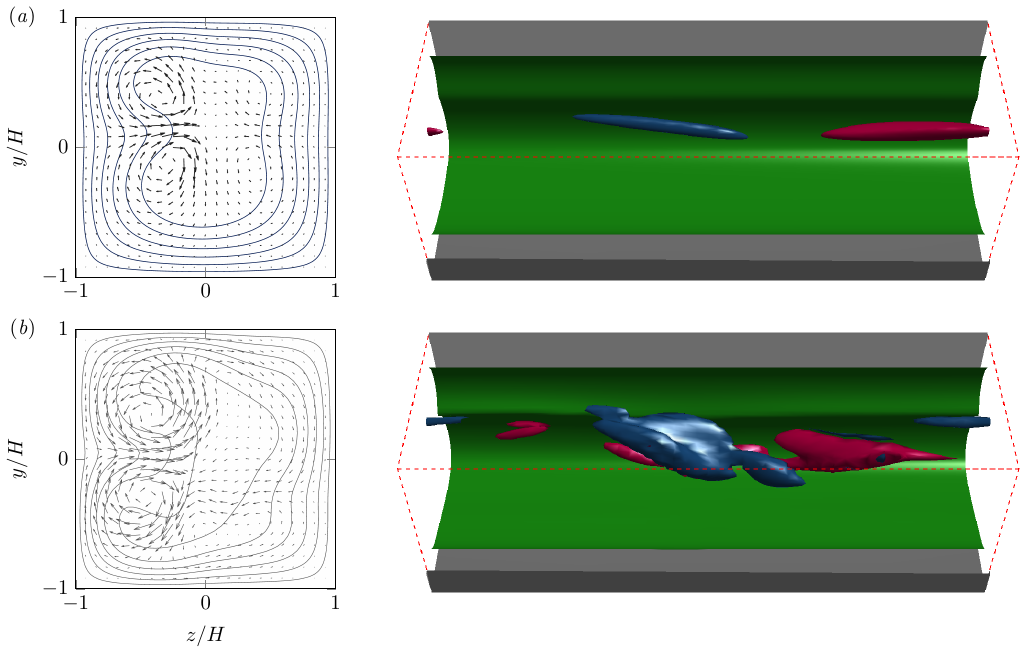}
    \caption
    {
    Comparison of instantaneous `two-vortex' states in
    (\textit{a}) the chaotic edge state \casenosymshort{2000} and
    (\textit{b}) the marginally turbulent state at $\Reb=1150$.
    Left column: Streamwise-averaged velocity fields $\xavg{\uvec}$ (iso-levels as in previous plots).
    The arrows in (\textit{b}) are scaled by a factor of two compared to those in (\textit{a}).
    Right column: Three-dimensional visualisation of the velocity field.
    The green iso-surface marks $u=0.75\ubulk$ for all $y\leq -z$.
    Vortices are identified as (\textit{a}) $Q^+ > 0.002$ and (\textit{b}) $Q^+ > 0.004$, respectively.
    Clockwise (red) and counter-clockwise rotation (blue) is measured by the local sign of $\omx$.
    To facilitate the comparison, the flow state in (\textit{b}) has been rotated by $\pi/2$ about the $x$-axis.
    }
\label{fig:uvw_xavg_3dview_EdgeTurbComp}
\end{figure}

%==============================================================
%
%
%
A closer look at the temporal variation of energy and dissipation is provided in \autoref{fig:Id_Einp_vs_Diss}, where the edge dynamics is presented as low-dimensional projections onto planes spanned by $\I$ and $\Diss$ as well as by $\Einp$ and $\Diss$. The data is contrasted with the joint probability functions (jpdf) for the corresponding quantities in the marginally turbulent reference simulation at $\Reb=1150$, for which selected snapshots have been presented in \autoref{fig:uvw_xavg_snaps_closedDuctref}.
The energy input-dissipation diagrams depicted in \autoref{fig:Id_Einp_vs_Diss}\textit{a-c} highlight that the edge states reside most of the time in the rather quiescent region of the state space in between the laminar equilibrium $(\Einplam,\Disslam)$ and what is the turbulent regime for $\Reb=1150$. For all edge states, the flow orbits along the cycles in a clockwise sense, implying that first the dissipation (or equivalently the enstrophy) experiences a steep rise owing to the strengthening of the quasi-streamwise vortices. After a small time lag of $\mathcal{O}(1\tbulk)$, the former low-speed streak turns into an intense high-speed region, causing the wall shear stress and thus the external energy input $\Einp$ to increase abruptly.
This is consistent with the different phases of the self-sustaining process discussed with the aid of the energy decomposition in equation~\eqref{eq:Epert_decomp} \citep{Waleffe_1997}, and are in line with the edge state dynamics
in plane Poiseuille flow \citep{Toh_Itano_2003} and
the asymptotic boundary layer \citep{Kreilos_Veble_Schneider_Eckhardt_2013}.

Also, \autoref{fig:Id_Einp_vs_Diss} provides further evidence for the symmetry break between $\Reb=1800$ and $\Reb=2000$ within the $\yrefl\zrefl$-symmetric edge state family:
Two distinct loops are discernible at the lower Reynolds number that collapse onto each other as the Reynolds number increases, reflecting that the full orbit is composed of two pre-periodic orbits that differ by a rotation of $\pi/2$ about the centreline.
During the bursting excursions, the chaotic edge states touch the lower end of the turbulent regime indicating the chaotic saddle, while their periodic counterparts in the symmetric subspaces reach much further into the turbulent `regime'. This is inasmuch expected as the imposed symmetries enforce turbulent activity to appear on two opposing walls for a subinterval of the cycle (in the $\yrefl$-symmetric subspace) or even for the full period (in the $\yrefl\zrefl$-symmetric subspace). The volume-averaged dissipation and energy-input should therefore attain higher values than in the symmetrically unconstrained case, where the vortical activity is -- most of the time -- of relevant amplitude near a single wall only.

For the sake of comparison, the parameter points associated with the `two-vortex' state and the `four-vortex' state presented in \autoref{fig:uvw_xavg_snaps_closedDuctref}\textit{a,b} are marked with white squares in \autoref{fig:Id_Einp_vs_Diss}\textit{a-c}, illustrating that the energy input and dissipation is in these case comparable to the values attained by some of the detected edge states.
Such `two-vortex' states, characterised by a strong concentration of the turbulent activity to a single wall, are in fact not that rare for flows under marginally turbulent conditions: For $20\%$ of the long-time turbulent trajectory at $\Reb=1150$, more than $50\%$ of the total streamwise-averaged enstrophy $\sum_i S_i$ resides in a single triangular sector.
The qualitative comparison between a `two-vortex' configuration of the chaotic edge state and one extracted from the chaotic marginally turbulent trajectory in \autoref{fig:uvw_xavg_3dview_EdgeTurbComp} highlights that the general flow configuration of a single low-velocity streak, surrounded by several counter-rotating quasi-streamwise vortices, along a single wall is comparable in both situations.
The difference lies in the amplitude of the vortices and thus the intensity of the generated secondary flow field, as well as in the streamwise length of the coherent structures: While the edge state features a single pair of streamwise elongated vortices, their counterparts in the turbulent state have a more complex shape and stretch less in the mean flow direction.

The fact that the turbulent trajectory transiently visits states in which a single wall/a pair of opposing walls dominates $\sum_i S_i$ is confirmed by the jpdfs presented in \autoref{fig:Id_Einp_vs_Diss}\textit{d-f}: The turbulent trajectory stays most of its lifetime at dissipation levels exceeding $1.6\Disslam$, featuring a vortical activity that is with $\abs{\I}\lesssim0.5$ rather uniformly distributed among the triangular sectors.
However, the presented data also indicate that the turbulent trajectory spends a non-negligible portion of its lifetime in a low-dissipation regime too. The latter is associated with a stronger focussing of the turbulent activity to a single wall/a pair of opposing walls, and the detected edge states are seen to oscillate between these two low-dissipation "tails" with indicator values $\abs{\I}>0.5$. The marginally turbulent trajectory, on the other hand, does not switch directly from a state with $\I\gtrsim0.5$ to one with $\I\lesssim0.5$ or vice versa, but it usually spends a (potentially long) while in the higher-dissipation region first (cf. also \citealp{Sekimoto_2011}).

Our findings thus imply that the low-dissipation excursions of a weakly turbulent trajectory to localised flow states are transient visits to one of the discussed edge states. The trajectories shadow the edge state dynamics for a certain time, potentially undergoing a bursting and wall-switching episode. Eventually, though, they get repelled from the edge along its single unstable direction back to the higher-dissipation regime of phase space, where flow states feature a significant vortical activity all over the cross-section.

\subsection{The edge state in state space}\label{sec:discuss_invsol}
In their seminal work on the edge state of plane Poiseuille flow, \citet{Toh_Itano_2003} observed periodic-like dynamics of alternating quiescent phases and energetic bursting episodes along the edge to turbulence, but their limited computational resources did not allow them to track the edge trajectory longer than one and a half cycles. So, \citet{Toh_Itano_2003} were raising the question whether the state space object they had identified was a periodic orbit or a heteroclinic cycle.
In contrast to a periodic orbit, a heteroclinic cycle consists of two saddles that are mutually connected in phase space by two heteroclinic orbits (cf. \citealp{Guckenheimer_Holmes_1983}, p. 46ff and \citealp{Wiggins_2003}, p. 631ff).

While \citet{Zammert_Eckhardt_2014} later re-analysed the edge for plane Poiseuille flow and confirmed that the edge state is indeed a periodic orbit, heteroclinic cycles were for a long time interpreted as the state space equivalent of turbulent bursts: \citet{Aubry_Holmes_Lumley_Stone_1988} found such an attracting heteroclinic cycle in a low-dimensional model of the near-wall region of a turbulent boundary layer. For most of their lifetime, trajectories were seen to stay near one of the two weakly unstable fix points -- representing relatively quiescent episodes of the turbulent activity. These episodes are interrupted when the trajectories eventually leave the fix points along their unstable manifolds, following the heteroclinic orbit to its sibling. In physical space, the dynamics along the heteroclinic orbits represent intense bursting events.
In a follow-up study, \citet{Armbruster_Guckenheimer_Holmes_1988} showed that it is in fact the $O(2)$-symmetry of the system that makes structurally stable heteroclinic cycles possible.
In the following years, similar statements were made for other systems obeying a $O(3)$-symmetry \citep{Guckenheimer_Holmes_1988} and even discrete $D_n$-symmetries \citep{Campbell_Holmes_1992}, underlining the high relevance of the system's symmetries for the existence of simple invariant sets in state space.

For the square duct, we have provided clear evidence that the symmetric edge states are limit cycles within the edge, and attempts to identify weakly unstable fixed points along the orbit were unsuccessful.
In this regard, our results are in good accordance with findings in $O(2)$-symmetric flow systems like
plane Poiseuille flows \citep{Zammert_Eckhardt_2014,Rawat_Cossu_Rincon_2014,%
Rawat_Cossu_Rincon_2016,Neelavara_Duguet_Lusseyran_2017} and
the asymptotic suction boundary layer \citep{Kreilos_Veble_Schneider_Eckhardt_2013,Khapko_al_2013}, whose edge states are all periodic orbits rather than heteroclinic cycles between two saddles.

Even though simple heteroclinic cycles could not be found within $\edgemfd$ for the discussed flows, the systems' symmetries still play a major role for the structure of the edge:
Topologically more complicated heteroclinic networks between pairs of symmetry-related saddles and stable fix points (here the edge states) turn out to occur in many systems and can be dynamically linked to periodic edge states.
So, \citet{Kreilos_Veble_Schneider_Eckhardt_2013} showed with a homotopy from plane Couette flow to the asymptotic suction boundary layer -- with the suction velocity at the wall as homotopy parameter -- that time-periodic edge states arise in a saddle–node infinite-period (SNIPER) bifurcation
(\citealp{Tuckerman_Barkley_1988}, \citealp{Tuckerman_2020}, \citealp{Strogatz_2018}, p.265f).
Under subcritical conditions, the edge features two symmetry-related edge states and two saddles separating them, all mutually connected by heteroclinic orbits. At the critical point, the stable fix points collide with the saddle points and give rise to a periodic orbit whose period is infinite at the bifurcation point, but decreases rapidly when moving away from it. Even for clearly supercritical conditions, the periodic orbit still features some properties of the previously present invariant structures in that the dynamics slow down significantly when the trajectory passes by the `ghosts' of the fix points in phase space.
A similar SNIPER bifurcation gives rise to the periodic edge states in plane Poiseuille flow under a variation of the lateral spatial domain size \citep{Rawat_Cossu_Rincon_2016}.
In the light of the qualitatively similar dynamics and taking into account the system's symmetries, we suspect that the periodic edge states in the square duct might also arise in a structurally similar SNIPER bifurcation.
In analogy to the situation in plane channel flow \citep{Rawat_Cossu_Rincon_2016}, a likely candidate for a parameter that drives the system into a SNIPER bifurcation is the cross-sectional aspect ratio $\AR=\Lz/\Ly$.
In order to check for the existence of such a bifurcation, though, additional efforts in form of a thorough homotopy study with $\AR$ as homotopy parameter are required.

%===================================================================
  \section{Conclusion}\label{sec:conclusion}
  In this study, we have investigated the dynamics of trajectories within the edge to square duct turbulence
for both the full state space and symmetric subspaces of the former.
Considering the full state space without symmetry constraints, the edge state is a chaotic attractor within $\edgemfd$ for the analysed parameter regime. The results are consistent with the chaotic edge states reported for circular pipe flows \citep{Duguet_Willis_Kerswell_2008}, which feature many similarities which the dynamics in the square duct.
In contrast to their counterparts in the pipe, though, the edge states in the square duct
experience phases of strong intermittent bursts that interrupt the otherwise rather
quiescent dynamics.
In physical space, these bursting episodes are seen to represent the different stages of the well-known
self-sustaining near-wall cycle \citep{Waleffe_1997,Hall_Sherwin_2010}:
The quiescent phases are characterised by a single straight low-speed streak that stretches along one of the four surrounding walls, flanked by counter-rotating quasi-streamwise vortices on either side.
The initially straight streak undergoes a linear instability and starts bending laterally,
while the vortices intensify and tilt w.r.t. the streamwise direction.
Eventually, the vortices lean over the low-speed streak, tear it apart and induce a pronounced high-speed
zone instead. Due to the particular geometry of the square duct, however, the intense vortices are located in the
corner regions of the cross-section and interact simultaneously with the near-wall region of both adjacent walls.
In other words: The large corner vortices simultaneously push high-momentum fluid towards one wall and induce a low-speed streak at the other. The new low-speed streak at the neighbouring wall slowly starts bending and the bursting cycle starts anew. With a length of $\mathcal{O}(100)\tbulk$ for both chaotic and time-periodic edge states, the characteristic time scale of the bursting events is much longer than that associated with their downstream propagation.

The outlined bursting cycles reveal a strong dynamical similarity with the periodic and chaotic edge states in
plane Poiseuille flow \citep{Zammert_Eckhardt_2014,Rawat_Cossu_Rincon_2014} and
the asymptotic boundary layer \citep{Khapko_al_2014,Kreilos_Veble_Schneider_Eckhardt_2013}.
But while the tilted vortices in these `planar' spanwise periodic flow configurations give birth to a new low-speed streak shifted by about half the spanwise period $\Lz/2$ to the left or right, here the vortices induce a new low-speed streak at a position rotated by $\pi/2$ or $3\pi/2$ about the duct centreline.
The structural similarity of the edge dynamics in both systems accentuates the role of the systems' symmetries that make such switching dynamics possible in the first place -- a continuous $O(2)$-symmetry in Poiseuille flow or the asymptotic boundary layer, and a discrete $D_n$-symmetry in the square duct.

In the `planar' spanwise-periodic cases, edge trajectories converge for a wide range of Reynolds numbers and domain sizes to periodic edge states that are limit cycles within $\edgemfd$. For the square duct, we detect such (relative) periodic edge states merely when restricting trajectories to distinct symmetric subspaces -- here specifically to states obeying a single or twofold mirror symmetry w.r.t. the duct's wall bisectors.
To the best of the authors' knowledge, the here reported edge states are the first periodic orbits found in a duct with rectangular cross-section.
The edge states in both symmetric subspaces feature very similar dynamics as their chaotic siblings in the unconstrained case -- with the difference that the imposed symmetries partly enforce bursting cycles to appear in symmetric pairs on two opposing walls.
A systematic variation of the Reynolds number confirmed the robustness of the detected period edge state dynamics over a significant range of Reynolds numbers, but it also revealed that the specific edge state characteristics change with the governing parameters in a highly non-uniform way -- in accordance with observations in `planar' spanwise-periodic shear flows \citep{Zammert_Eckhardt_2014}. While the edge states' period varies almost linearly with $\Reb$ in some parameter regimes, sudden steep rises of the period in terms of a period doubling or a symmetry break are observed elsewhere in the parameter space.

The current results indicate that the localisation of a single turbulent regeneration cycle to one or two of the surrounding walls is a fundamental feature of the edge state dynamics in square duct flow. Also, it was revealed that marginally turbulent trajectories regularly attain similar low-dissipation states, in which turbulent perturbations are concentrated near one or two walls only. Together, these observations raise the question whether such `two-vortex' and 'four-vortex' episodes along marginally turbulent trajectories can be associated with transient visits to the edge states detected in the course of this study. Within this dynamical systems perspective, the trajectories are expected to approach an edge state via its stable manifold that locally collapses with $\edgemfd$, to shadow its dynamics for a certain time interval and to eventually get repelled along the single unstable direction towards the turbulent saddle/attractor.
In order to check the validity of this hypothesis, further efforts are necessary to identify such excursions to the vicinity of an edge state along a marginally turbulent trajectory and to verify that these are indeed accompanied by a localisation of the turbulent activity to one or two walls.

More generally, it remains to clarify how the edge state dynamics change when the cross-sectional aspect ratio $\AR$ is increased towards rectangular ducts of non-square cross-section. Although some preliminary results prove the general existence of spanwise-localised travelling wave solutions in the rectangular case $\AR>1$ \citep{Okino_2011}, a comprehensive investigation of the edge dynamics that could clarify how the intermittent bursting dynamics and the wall-switching behaviour change with $\AR$ is lacking.
With the observation that periodic edge states can arise in the context of SNIPER bifurcations in mind
\citep{Kreilos_Veble_Schneider_Eckhardt_2013,Rawat_Cossu_Rincon_2016}, a comparison of the edge state dynamics at successively increasing aspect ratio will allow to verify whether or not the here identified periodic edge states are born in a similar global bifurcation. 
Recalling that plane Poiseuille flow represents the asymptotic state of an infinitely wide duct as $\AR\to\infty$, we expect such a continuation in the duct's aspect ratio to moreover provide valuable information on how edge states in the rectangular duct are dynamically connected to those in plane channel flow.

%===================================================================
  \section*{Funding}
  This work was supported in part by the European Research Council under the Caust grant ERC-AdG-101018287. In this context, we want to express our thanks to Prof J. Jim{\'e}nez and Dr K. Osawa for their kind hospitality during the fifth Madrid turbulence summer workshop. We also thank Prof M. Avila, Dr D. Mor{\'o}n and Prof L. van Veen for very helpful discussions.
Part of the simulations were performed on the supercomputer bwUniCluster funded by the Ministry of Science, Research and the Arts Baden-W{\"u}rttemberg and the Universities of the State of Baden-W{\"u}rttemberg. The computer resources, technical expertise and assistance provided by the staff are gratefully acknowledged.

%===================================================================
  \vspace*{2ex}
  \section*{Declaration of Interests}
  The authors report no conflict of interest.

  \vspace*{2ex}
  \section*{ORCID}
  M. Scherer,
  \href{https://orcid.org/0000-0002-6301-4704}{https://orcid.org/0000-0002-6301-4704};\\
  M. Uhlmann,
  \href{https://orcid.org/0000-0001-7960-092X}{https://orcid.org/0000-0001-7960-092X};\\
  G. Kawahara,
  \href{https://orcid.org/0000-0001-7414-0477}{https://orcid.org/0000-0001-7414-0477}.

%===================================================================
  \appendix
  %\section{Turbulent reference cases}
%
%\input{sections/tables/tableRefCases.tex}
%

\section{Convergence to a periodic edge state}\label{app:appendix_conv}

%
%==============================================================
% insert figure:
\begin{figure}%[tp]
    \centering
    \includegraphics[width=0.75\linewidth]
    {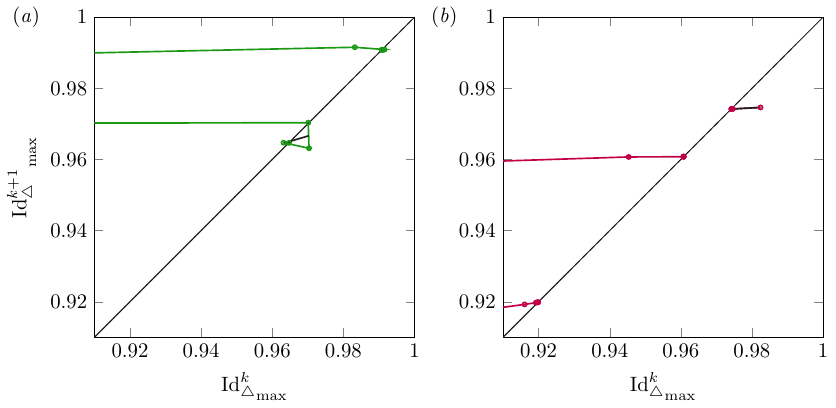}
    \caption
    {
    First-return map of the indicator function $\I$ in terms of the peak values of the signal
    for selected periodic orbits in the
    (\textit{a}) $\yrefl$- and (\textit{b}) $\yrefl\zrefl$-symmetric subspace. % ($\Reb\in\{2000,2600,3200\}$).
    Thin black lines represent the best fit linear approximation to the map
    $\Id_{\max}^{k+1} = \mathrm{P}(\Id_{\max}^{k})$,
    computed for the last three points in the sequence.
    }
\label{fig:PoincareMap_Idmax_PO_yrefl_yzrefl}
\end{figure}

%==============================================================
%
%
%
The time signal $\I(t)$ is used to estimate the fundamental period of the shadowed limit cycle, since it contains more information on the dynamics than the `pure' energy signal $\Epertrms$. In particular, it distinguishes different states that are identical modulo finite rotations.
To check for convergence, an integral $L_2$-error between two consecutive cycles $\I^{k-1}$ and $\I^{k}$ of the expected period length $\tper$ is introduced as
\begin{equation}
  \varepsilon = \dfrac{\Lnorm{\I^{k}-\I^{k-1}}{2,T}}{\Lnorm{\I^{k}}{2,T}},
  \label{eq:L2norm_conv_criterion}
\end{equation}
where the $L_2$-norm over a full (guessed) period is defined as
\begin{equation}
  \Lnorm{\bullet}{2,T}^2 = \displaystyle\int\limits_{t'=t}^{t+T}\; \bullet^{2}  \;\mathrm{d}t'.
\end{equation}
As period guess $T$, we choose the time at which the temporal auto-correlation function of $\I(t)$ (excluding an initial transient) attains its non-trivial global maximum. An edge trajectory is then considered converged to a periodic orbit if the error $\varepsilon$ defined in equation~\eqref{eq:L2norm_conv_criterion} falls below $5\%$.
The only exception is case~\caseyrefl{3200}, for which the two pre-periodic orbits feature the same $\I(t)$ profile.
In this case, we quantify convergence instead based on the signal of the streamwise-averaged enstrophy in the single sectors, $S_i(t)$.

\section{Stability of the periodic edge state}\label{app:appendix_stab}
That the shadowed periodic orbits are indeed limit cycles is shown in \autoref{fig:PoincareMap_Idmax_PO_yrefl_yzrefl}, where a first-return map is presented for the peaks of the signal $\I(t)$ for selected periodic orbits in both considered symmetric subspaces \citep{Khapko_al_2013,Zammert_Eckhardt_2014}.
Following classical dynamical systems theory (\citealp{Guckenheimer_Holmes_1983}, pp.~22), this low-dimensional projection of the full system is strictly periodic if its Poincar{\'e} map converges to a fix point. This is the case if the sequence in the direct neighbourhood of the fix point follows a straight line with slope $|\gamma| < 1$.
The black lines in \autoref{fig:PoincareMap_Idmax_PO_yrefl_yzrefl} represent the linear trend for the last few iterations and verify that this condition is indeed fulfilled for the selected cases.

%===================================================================
  \phantomsection
  \bibliographystyle{plainnatdoilinknourl}
  \setlength{\bibsep}{.4ex}
  \addcontentsline{toc}{section}{References}
  \bibliography{%
    literature_invsol,%
    literature_numerics,%
    literature_duct,%
    literature_extremeEvents
  }

%===================================================================

\end{document}